\documentclass[prb,twocolumn]{revtex4}
\usepackage{graphicx}
\usepackage{hyperref}
\usepackage{amssymb}
\usepackage{dcolumn}
\usepackage{float}
\usepackage{bm}
\usepackage{booktabs}
\usepackage{amsmath} 
\usepackage[utf8]{inputenc}
\usepackage{lmodern}
\usepackage{color}
\definecolor{red}{rgb}{1.0,0.0,0.0}
\usepackage{makecell}
\usepackage{marginnote}

\DeclareMathAlphabet{\bi}{OML}{cmm}{b}{it}

\def\ba{\begin{aligned}}
\def\ea{\end{aligned}}
\def\be{\begin{equation}}
\def\ee{\end{equation}}
\def\bearr{\begin{eqnarray}}
\def\eearr{\end{eqnarray}}

\def\l{\left}
\def\r{\right}

\usepackage{tikz}

\begin{document}
\title{Berry curvature induced magnetotransport in 3D noncentrosymmetric metals}
\bigskip
\author{Ojasvi Pal, Bashab Dey and Tarun Kanti Ghosh\\
\normalsize
Department of Physics, Indian Institute of Technology-Kanpur,
Kanpur-208 016, India}
\begin{abstract}
We study the magnetoelectric and magnetothermal transport properties of
noncentrosymmetric metals using semiclassical Boltzmann transport
formalism by incorporating the effects of Berry curvature and orbital
magnetic moment. These effects impart quadratic-$B$ dependence to the
magnetoelectric and magnetothermal conductivities, leading to intriguing
phenomena such as planar Hall effect, negative magnetoresistance, planar
Nernst effect and negative Seebeck effect. The transport coefficients
associated with these effects show the usual oscillatory  behavior  with
respect  to  the  angle  between  the  applied  electric  field  and
magnetic field. The bands of noncentrosymmetric  metals  are  split  by
Rashba  spin-orbit  coupling  except  at a  band  touching point. For Fermi energy below (above) the band touching point,
giant (diminished) negative magnetoresistance is observed. This difference in the nature of magnetoresistance is related to the magnitudes of the velocities, Berry curvature and orbital
magnetic moment on the respective Fermi surfaces, where the orbital magnetic moment plays the dominant role. The absolute
magnetoresistance and planar Hall conductivity show a decreasing
(increasing)  trend with Rashba coupling parameter for Fermi energy below
(above) the band touching point.  
\end{abstract}
\maketitle
\section{Introduction}
The interaction of electron's spin with its motion, known as Spin-Orbit Coupling (SOC), has an extensive role
in condensed matter physics. Spintronics\cite{spintronics1,spintronics2,spintronics3,spintronics4} emerges as a multidisciplinary field dealing with the active manipulation of spin degree of freedom as spin is a non-conserved quantity in spin-orbit coupled systems. The semiconductor heterojunction undergoes the breaking of inversion symmetry due to the interfacial electric field giving rise to Rashba spin-orbit interaction (RSOI)\cite{rashba1,rashba2}. The recent developments that capture RSOI across various fields of physics and material science includes spin Hall effect\cite{she1,she2,she3,she4,she5,she6}, topological insulator\cite{TI}, spin-orbit torque\cite{she6,sot}, and spin galvanic effects\cite{she6,sge} - all covering under the umbrella of emerging spin-orbitronics, a branch of spintronics that centers around the control of non-equilibrium material properties utilizing SOC\cite{spinorbitronics}.

The scarcity of semiconductors having a large RSOI hinders the growth of spintronic devices in actual practice. Later, various theoretical and experimental investigations suggested that the systems like Bi/Ag(111)\cite{Bi/Ag} surface alloy, Bi$_{2}$Se$_{3}$\cite{Bi2se31,Bi2se32}, and 3D polar semiconductor BiTeX (X=Cl, Br, I) show stronger spin-orbit coupling than conventional semiconductor heterostructures, which enhances the spintronics applications. The potential candidates in this category are bismuth tellurohalides having the trilayer structure with  X-Bi-Te stacking. The giant RSOI\cite{giant1,giant2,giant3,giant4,giant5} appears in BiTeX compounds  due to structural asymmetry. Apart from BiTeX compounds, B20 compounds\cite{B20} and noncentrosymmetric metals like {Li}$_{2}$(Pd$_{1-x}$Pt$_{x})_{3}$B\cite{LiPd} show large RSOI ($\alpha\sim(0.1-3)\times {10}^{-10}$ eV-m )\cite{zhou} as a result of inversion symmetry (IS) breaking. These  compounds have cubic crystal structure and the symmetries of these materials are analyzed in order to construct the effective low-energy Hamiltonian. $B20$ compounds belong to the space group  $P_{{2_1}3}$, where subscripts 2 and 3 represents the two-fold and three-fold rotational symmetries respectively and subscript 1 shows a fractal translation in the space-group operation. For $P_{{2_1}3}$, the $K$ group which contains all the point-group operations within the space group is the $T23$ group\cite{B20}. The point symmetry of {Li}$_{2}$(Pd$_{1-x}$Pt$_{x})_{3}$B is described by the cubic point group $G=O$\cite{symmetryLiPd,symmetry1}. The broken IS also results in nontrivial Berry curvature (BC)\cite{berry} in the system. The electrical and thermoelectric transport properties of the system in the absence of magnetic field are not affected by BC and has been studied recently\cite{sonu}. However, in the presence of weak magnetic field, the BC and orbital magnetic moment (OMM) produces significant modification in the transport properties which is the prime focus of our paper.

The BC acts as a magnetic field in the momentum space\cite{xiao} and leads to various interesting phenomena that are typically absent in conventional condensed matter systems. Some examples are anomalous Hall effect\cite{AHE1,xiao,AHE2,kamal} and anomalous Nernst effect\cite{ANE1,ANE2,ANE3} which are predicted to exist in systems with broken time reversal symmetry. The presence of BC may also result in Chiral magnetic effect (CME) in which equilibrium current is generated by magnetic field without any external bias in the presence of finite chemical potential\cite{CME1,CME2}. The existence of external parallel electric and magnetic field results in semiclassical manifestation of the effect known as chiral anomaly\cite{chiral1,chiral2,chiral3,CME2}. It is a purely quantum mechanical effect describing the anomalous nonconservation of a chiral current. 

To address the negative magnetoresistance (MR) in Weyl metal, a concept of weak anti-localization quantum corrections in the collision term and BC through the semiclassical equations of motion was included in the Boltzmann transport framework\cite{sasaki}. In topological semimetals, a negative MR\cite{1,2,3,4,5,6,7,8,9,sasaki} is explained as an effect of chiral anomaly. But in some systems like topological insulators, e.g., Bi$_{2}$Se$_{3}$, a chiral anomaly is not well-defined. In such cases, the negative MR is explained by anomalous velocity induced by BC and OMM\cite{lu}. Using semiclassical equations of motion, one can observe that velocity contains an anomalous term due to the presence of BC which is proportional to $\mathbf{B}$ in the linear-response limit and in turn, a correction to the conductivity becomes proportional to $B^{2}$. In this way, the BC results in negative MR while the OMM enhances it by modifying the band velocity through a Zeeman-like term in the energy dispersion.  

One of the striking consequences of BC is the planar Hall effect (PHE)\cite{nandy, burkov}. The PHE appears in a configuration when the applied electric field, magnetic field and the induced voltage lie in the same plane such that the induced voltage is perpendicular to the electric field.  When the electric field is applied along $\hat{x}$ direction and the magnetic field is applied in the $x$-$y$ plane making an angle with the $x$-axis, then the transverse conductivity measured along $\hat{y}$ defines the planar Hall conductivity (PHC).

The BC driven transport also produces the Nernst effect which describes the transverse electric response to a thermal gradient. In conventional Nernst effect, the induced voltage, temperature gradient and magnetic field are mutually perpendicular to each other. An anomalous Nernst effect demands the need of non-zero Berry curvature component parallel to the ($\hat{\mathbf{E}} \times \boldsymbol{\nabla}{T}$) plane. The planar Nernst effect (PNE) occurs in a configuration where the temperature gradient, magnetic field and voltage are co-planar such that the induced voltage is perpendicular to the temperature gradient and it is equivalent to the transverse thermopower\cite{ANE3,kamal2}. A similar kind of phenomenon is known to exist in ferromagnetic systems\cite{pne1 ferro,pne2 ferro}.

Motivated by the unconventional phenomena produced due to BC, we calculate various magnetotransport coefficients such as electrical conductivity, PHC, MR, Seebeck coefficient (SC) and planar Nernst coefficient (PNC) of noncentrosymmetric metals. We have explicitly included OMM in our calculations which arises due to the self-rotation of wave packet of the Bloch electron.\cite{niu2}.

This work is organized as follows: In Sec. \ref{II}, we present a discussion on the low-energy band structure of noncentrosymmetric metals. In Sec. \ref{III}, we provide a review of the semiclassical Boltzmann transport formalism incorporating the BC and OMM  effects. We provide a general expressions of the magnetoconductivities in Sec. \ref{IV}(A) and discuss the result of PHE and MR in Sec. \ref{IV}(B). It is followed by the formalism of thermoelectric transport and general expression of thermoelectric coefficients in Sec. \ref{V}(A). The results of PNE and SC are given in Sec. \ref{V}(B). Finally, we conclude and summarize our main results in Sec. \ref{VI}.
\section{Low-energy band structure}\label{II}
The noncentrosymmetric metals such as {Li}$_{2}$(Pd$_{1-x}$Pt$_{x})_{3}$B and B20 compounds exhibit cubic crystal structure. In such lattice geometry, the effective low-energy Hamiltonian of conduction electrons based on symmetry analysis\cite{symmetry1,B20} is given by:
\begin{equation}\label{hamiltonian}
H = \frac{{\hbar}^2 {\mathbf{k}}^2}{2{m^\ast}}{ \sigma_0} + \alpha  {\boldsymbol{\sigma}\cdot {\mathbf{k}}}.
\end{equation} 
Here, ${m^\ast}$ is the effective mass of an electron, $\boldsymbol{\sigma} = ( \sigma_{x},\sigma_{y},\sigma_{z})$ is the vector consisting of the three Pauli matrices, $\sigma_{0}$ is $2\times2$ identity matrix, ${\mathbf{k}}$ is the electron wave vector with magnitude, $k = \sqrt{{k_x}^2+{k_y}^2+{k_z}^2}$. The term  $\alpha$ denotes the strength of Rashba spin-orbit interaction (RSOI). The Hamiltonian in Eq. (\ref{hamiltonian}) preserves the time-reversal symmetry (TRS) and breaks the inversion symmetry. 
\begin{figure}[htbp]
\hspace{0cm}\includegraphics[trim={0cm 1.5cm 0cm  0cm},clip,width=9cm]{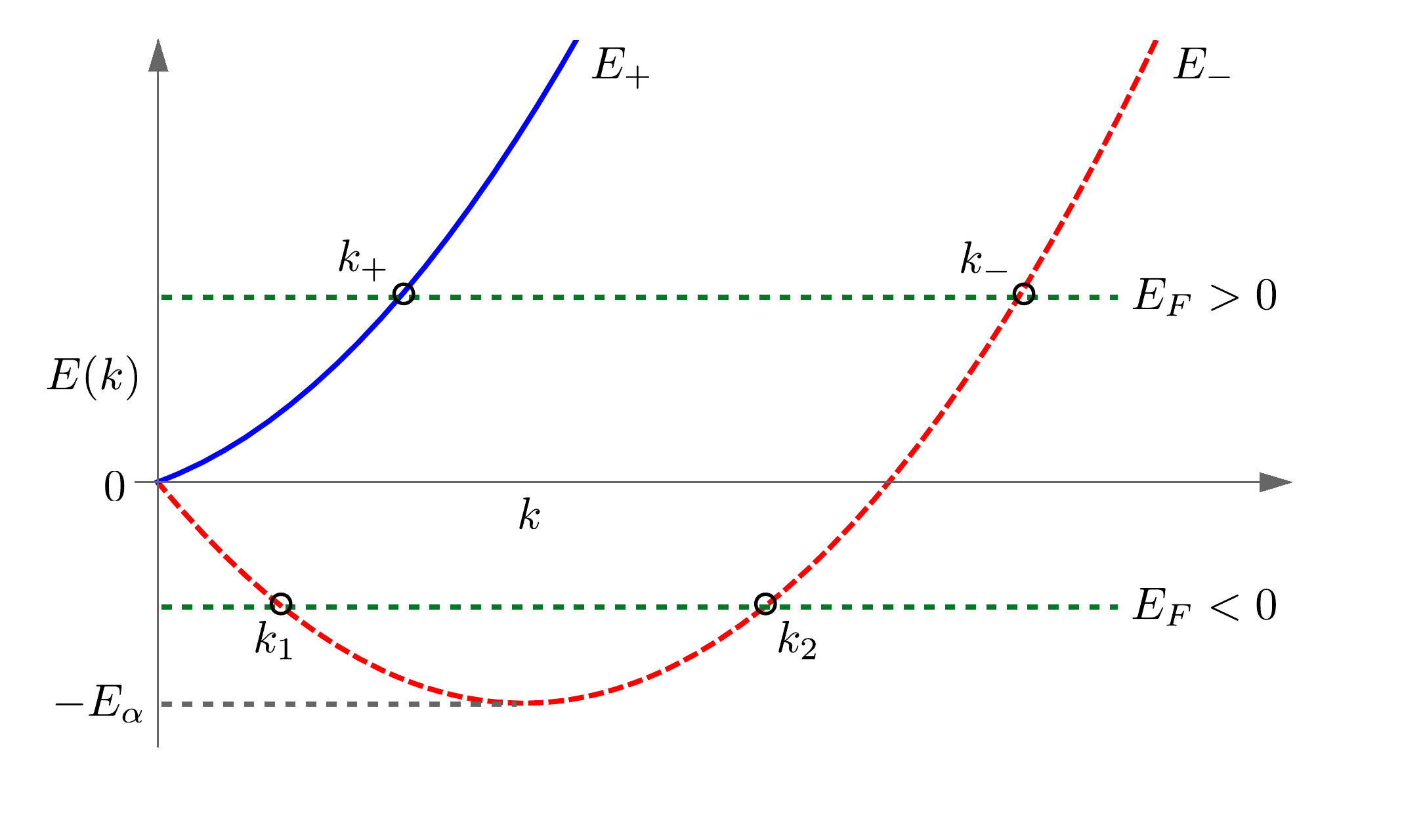}
\caption{The sketch of low-energy band structure of noncentrosymmetric metals: The two bands touch at $k=$0 which is known as band touching point (BTP). The Fermi surface intersect both the bands when ${E_F} >$ 0 and intersects only the lower band at two Fermi wave vectors for ${E_F} <$ 0.}
\label{plot-ncm}
\end{figure}The energy dispersion is given by
\begin{equation}
{E_{k\lambda}} = \frac{{\hbar}^2 {k}^2}{2{m^\ast}} + \lambda \alpha k,
\end{equation}
where $\lambda$ = $\pm$ representing the two spin-split bands and their corresponding eigen states as $ {\psi}_{\mathbf{k}}^{\lambda}(\mathbf{r})= {{u}^{\lambda}_{\mathbf{k}}}{ e^{\iota\mathbf{k}\cdot\mathbf{r}}}/{\sqrt{V}}$ with\\
\begin{equation}
|u_{\bf k}^+\rangle = 
 \left (\begin{matrix}
  {\cos{(\theta/2)} } \\
{ e^{\iota\phi}\sin{(\theta/2)}} 
\end{matrix} \right),\\
\end{equation}
\\
\begin{equation}
 |u_{\bf k}^-\rangle =  \left (\begin{matrix}
  {\sin{(\theta/2)} } \\
{- e^{\iota\phi}\cos{(\theta/2)}} 
\end{matrix} \right).
\end{equation} 

The energy dispersion is depicted in Fig. \ref{plot-ncm}. At $k=0$, the two bands touch each other which is known as band touching point (BTP). The wave vectors corresponding to $E>0$ are given by ${k_\lambda}=-\lambda{k_\alpha}+\sqrt{{{k_\alpha}^2}+2{m^\ast}E/{\hbar^2}}$, where $ {k_\alpha}={{m^\ast}\alpha}/{{\hbar}^2}$ is the wave vector associated with the SOI and the corresponding energy scale is ${E_\alpha}={\hbar}^2 {k_\alpha}^2/{2{m^\ast}}$ and ${k_{\pm}}$ refers to the radii of two concentric spherical constant energy surfaces. It consists of both $\lambda=-$ and $\lambda=+$ bands  and therefore contributes to the following density of states (DOS):
\begin{equation}
D_{\lambda}^{>}(E)=\frac{1}{4{\pi^2}}{\bigg({\frac{2{m^\ast}}{\hbar^2}}\bigg)^{3/2}}\bigg[\frac{E+2{E_\alpha}}{\sqrt{E+{E_\alpha}}}-\lambda\sqrt{4{E_\alpha}}\bigg].
\end{equation}
For $E < 0$, only the energy band corresponding to $\lambda = -$ exists with minima located at $k={k_\alpha}$ having $E = {E_{min}} = -{E_\alpha}$. When energy lies in the range $-{E_\alpha}<E<0$, the energy surface will dissect the $E_-$ band into two spherical shells of radii, ${k_\eta}={k_\alpha}-(-1)^{\eta-1}\sqrt{{{k_\alpha}^2}+2{m^\ast}E/{\hbar^2}}$, where $\eta=1, 2$ is the branch index. The region between these two concentric spherical shells $({0 \leq {k_1} \leq {k_\alpha}}$ and ${{k_\alpha} \leq {k_2} \leq 2{k_\alpha}}) $ contains the given DOS:
\begin{equation}
D_{\eta}^{<}(E)=\frac{1}{4{\pi^2}}{\bigg({\frac{2{m^\ast}}{\hbar^2}}\bigg)^{3/2}}\bigg[\frac{E+2{E_\alpha}}{\sqrt{E+{E_\alpha}}}-(-1)^{\eta-1}\sqrt{4{E_\alpha}}\bigg].
\end{equation}
The BC can be calculated using $\bm{\Omega}_{\mathbf{k}}^{\lambda} = i{ \boldsymbol{\nabla}}_{\mathbf{k}}\times (\langle u^{\lambda}_{\mathbf{k}}|  \boldsymbol{\nabla}_\mathbf{k} |u^{\lambda}_{\mathbf{k}}\rangle) $ as\cite{xiao} 
\begin{equation}
\bm{\Omega}_{\mathbf{k}}^{\lambda} = -\frac{\lambda{\mathbf{k}}}{2{k^3}}.
\end{equation}
The orbital magnetic moment (OMM) arises from the self-rotation of a wave packet formed by superposing the Bloch states of a band. The OMM is calculated using $\mathbf{m}_{\mathbf{k}}^{\lambda} = {-i e}/{2\hbar} \langle \boldsymbol{\nabla}_\mathbf{k}{u}^{\lambda}_{\mathbf{k}}| \times \l(H({\mathbf{k}}) - E({\mathbf{k}})\r) | \boldsymbol{\nabla}_\mathbf{k}{u}^{\lambda}_{\mathbf{k}}\rangle$ to be\cite{OMM1,OMM2}
\begin{equation}  
\mathbf{m}_\mathbf{k}^{\lambda} = \frac{\lambda {\alpha}ek}{\hbar}\bm{\Omega}_{\mathbf{k}}^{\lambda} = -\frac{\alpha e{\mathbf{k}}}{2\hbar{k^2}}.
\end{equation}
The system has time-reversal invariance and broken spatial inversion symmetry, thus the BC satisfies  $\bm{\Omega}^{\lambda}(-{\mathbf{k}})=-\bm{\Omega}^{\lambda}({\mathbf{k}})$. The OMM transforms exactly like BC under symmetry operations. 
\section{Semiclassical transport formalism}\label{III}
We use Boltzmann transport formalism to understand the effects of Berry curvature and OMM on the electrical and thermoelectric transport  properties of the system. The semiclassical Boltzmann approach works effectively under the regime of low magnetic field and therefore, the formation of Landau levels is irrelevant in our case.

The modified semiclassical equations of motion of a Bloch wave packet (including the effects of BC and OMM) are given by\cite{xiao,moore}
\begin{equation}{\label{EOM1}}
\mathbf{\dot r}^{\lambda}=\frac{1}{{D}_{\mathbf{k}}^{\lambda}}\bigg[\tilde{\mathbf{v}}_{\mathbf{k}}^{\lambda}+\frac{e}{\hbar}(\mathbf{E} \times \bm{\Omega}_{\mathbf{k}}^{\lambda})+ \frac{e}{\hbar}(\tilde{\mathbf{v}}_{\mathbf{k}}^{\lambda} \cdot \boldsymbol{\Omega}_{\mathbf{k}}^{\lambda})\mathbf{B}\bigg],
\end{equation}
\begin{equation}{\label{EOM2}}
\hbar \mathbf{\dot k}^{\lambda} = \frac{1}{{D}_{\mathbf{k}}^{\lambda}}\bigg[-e \mathbf{E}- e(\tilde{\mathbf{v}}_{\mathbf{k}}^{\lambda} \times \mathbf{B})- \frac{e^2}{\hbar}(\mathbf{E} \cdot \mathbf{B}) \boldsymbol{\Omega}_{\mathbf{k}}^{\lambda}\bigg],
\end{equation}
where we have defined ${D}_{\mathbf{k}}^{\lambda} = ({1+\frac{e\mathbf{B} \cdot \bm{\Omega}_{\mathbf{k} }^{\lambda}}{\hbar}})$. The phase-space volume is changed by a factor of $[D(k)]^{-1}$\cite{mdos}. Density of states no longer remains constant for Berry phase modified dynamics. In order to balance this changed phase-space volume, density of states is multiplied by $[D(k)]$ such that the number of states in the volume element remains constant in time. 

The semiclassical band velocity is $\hbar\tilde{\mathbf{v}}_{\mathbf{k}}^{\lambda} =  \boldsymbol{\nabla}_{\mathbf{k}}{\tilde{E}}_{\mathbf{k}}^{\lambda}$, where ${\tilde{E}}_{\mathbf{k}}^{\lambda}$ = ${{E}}_{\mathbf{k}}^{\lambda}- \mathbf{m}_{\mathbf{k}}^{\lambda}\cdot \mathbf{B}$ is the modified energy including Zeeman-like coupling of OMM with external magnetic field. The OMM modified velocity is expressed as $\tilde{\mathbf{v}}_{\mathbf{k}}^{\lambda} = {\mathbf{v}}_{\mathbf{k}}^{\lambda} + {\mathbf{v}}_{\mathbf{k}}^{\textbf{m},{\lambda}}$, 
where ${\mathbf{v}}_{\mathbf{k}}^{\lambda}$ is proportional to $\sqrt{\frac{2({E}^{\lambda}+ E_{\alpha})}{m}}$ and 
$\hbar{\mathbf{v}}_{\mathbf{k}}^{\textbf{m},{\lambda}} = -{ \boldsymbol{\nabla}}_{\mathbf{k}}(\mathbf{m}_{\mathbf{k}}^{\lambda}\cdot \mathbf{B})$, which is dependent on the orientation of  external magnetic field. 

It is to be noted that we have considered the energy correction only upto linear order in $B$. Since $\mathbf{m_{k}}$ is band independent, the energy correction is identical for both the bands. Hence, the term representing the coupling is added to the Hamiltonian as a constant diagonal matrix given by $-(\mathbf{m_{k}}\cdot \mathbf{B}){\sigma_{0}}$ which does not alter the form of eigenfunctions. So, the Berry curvature does not acquire any modification from the magnetic field due to OMM coupling. The second-order energy corrections are band dependent which may modify the eigenfunctions and BC to give the equal order of contribution in the conductivities\cite{prlniu,potential,shift}.

Equation (\ref{EOM1}) holds two important effects: the term $(\mathbf{E} \times \bm{\Omega}_{\mathbf{k}}^{\lambda})$ is responsible for the anomalous Hall effect (AHE)\cite{AHE1,xiao,AHE2} and the third term $(\tilde{\mathbf{v}}_{\mathbf{k}}^{\lambda} \cdot \boldsymbol{\Omega}_{\mathbf{k}}^{\lambda})\mathbf{B}$ give rise to the chiral magnetic effect\cite{CME1,CME2}. In Eq. (\ref{EOM2}), the first two terms describes the Lorentz force whereas the last term $(\mathbf{E} \cdot \mathbf{B}) \boldsymbol{\Omega}_{\mathbf{k}}^{\lambda}$ denotes the chiral anomaly effect. 

The steady-state Boltzmann transport equation (BTE) to obtain the non-equilibrium distribution function $\tilde{f}_{\mathbf{r},\mathbf{k}}^{\lambda}$ is given as\cite{ashcroft}
\begin{equation}
\mathbf{\dot r}^{\lambda} \cdot { \boldsymbol{\nabla}}_{\mathbf{r}} \tilde{f}_{\mathbf{r},\mathbf{k}}^{\lambda} + \mathbf{\dot k}^{\lambda} \cdot{ \boldsymbol{\nabla}}_{\mathbf{k}} \tilde{f}_{\mathbf{r},\mathbf{k}}^{\lambda}= {I}_{\textnormal{coll}} \{{\tilde{f}_{\mathbf{r},\mathbf{k}}^{\lambda}}\}. 
\end{equation}
The collision integral under the relaxation time approximation can be written as
\begin{equation}
{I}_{\textnormal{coll}} \{{\tilde{f}_{\mathbf{r},\mathbf{k}}^{\lambda}}\} = -\frac{\tilde{f}_{\mathbf{r},\mathbf{k}}^{\lambda}-\tilde{f}_{\textnormal{eq}}^{\lambda}}{{\tau}_{\mathbf{k}}^{\lambda}},
\end{equation}\\
where ${\tilde{f}}_{\textnormal{eq}}^{\lambda} = {[1+ e^{\beta(\tilde{E}_{\mathbf{k}}^{\lambda}-\mu)}]}^{-1}$ is the Fermi-Dirac distribution Function and ${\beta}^{-1}\equiv{k_B}T$, $\mu$ denotes  the thermal energy and chemical potential respectively. The quantity ${\tau}_{\mathbf{k}}^{\lambda}$ is the relaxation time. We consider the relaxation time to be constant in our case. Thus, the BTE becomes
\begin{equation}
\mathbf{\dot r}^{\lambda} \cdot\nabla_{\mathbf{r}} \tilde{f}_{{\mathbf{r}},{\mathbf{k}}}^{\lambda} + \mathbf{\dot k}^{\lambda} \cdot\nabla_{\mathbf{k}}\tilde{f}_{{\mathbf{r}},{\mathbf{k}}}^{\lambda} =-\frac{\tilde{f}_{{\mathbf{r}},{\mathbf{k}}}^{\lambda}-\tilde{f}_{\textnormal{eq}}^{\lambda}}{\tau}.
 \end{equation}
Within the regime of linear response theory, the non-equilibrium distribution function (NDF) is given as
\begin{equation}{\label{NDF}}
\begin{aligned}
\tilde{f}_{{\mathbf{r}},{\mathbf{k}}}^{\lambda}=&{\tilde{f}}_{\textnormal{eq}}^{\lambda}+\bigg[\frac{\tau}{{D}_{\mathbf{k}}^{\lambda}}
\left(-e \mathbf{E}-\frac{(\tilde{E}_{\mathbf{k}}^{\lambda}-\mu)}{T}\boldsymbol{\nabla}_{\mathbf{r}}{T}\right)\\
&\cdot \bigg( \tilde{\mathbf{v}}_{\mathbf{k}}^{\lambda}+\frac{e}{\hbar}{\mathbf{B}}
({\tilde{\mathbf{v}}_{\mathbf{k}}^{\lambda}}\cdot \bm{\Omega}_{\mathbf{k}}^{\lambda}) \bigg)+ \tilde{\mathbf{v}}_{\mathbf{k}}^{\lambda}\cdot \mathbf{\Gamma}^{\lambda} \bigg]\bigg(-\frac{{\partial{\tilde{f}_{\textnormal{eq}}^{\lambda}}}}{\partial{{\tilde{E}}_{\mathbf{k}}^{\lambda}}} \bigg).
\end{aligned}
\end{equation}
The second term represents the deviation due to the scattering process and it contributes to the non-equilibrium part of current induced by the electric field and statistical forces like temperature gradient. The term $(\tilde{\mathbf{v}}_{\mathbf{k}}^{\lambda}\cdot \mathbf{\Gamma}^{\lambda})$ in the obtained NDF shows the contribution of the Lorentz force\cite{jacobini,sasaki}. The main motivation behind our paper is to study the BC induced magnetotransport phenomena and therefore, we have not included the Lorentz force contribution.

The electric current upto first order in electric field and gradient in temperature is defined as\cite{ashcroft}
 \begin{equation}\label{j_i}
 {j}_{i}^{e}=\sigma_{ij}{E_j}+\alpha_{ij}(-\nabla_{j}T),
 \end{equation}
where $\sigma_{ij}$ and $\alpha_{ij}$ represents the elements of electrical conductivity matrix and thermoelectric conductivity matrix respectively.
The current density is defined as 
\begin{equation}{\label{j_electric}}
{\mathbf{j}} = -e\sum_{\lambda=\pm1}\int[d\mathbf{k}]{D}_{\mathbf{k}}^{\lambda}({\dot{{\mathbf{r}}}^{\lambda}})\tilde{f}_{{\mathbf{r}},{\mathbf{k}}}^{\lambda},
\end{equation}
where $[d\mathbf{k}]={d^3}k/{(2\pi)}^3$. But quantum mechanically, the carrier is represented by the wave packet which has finite spread in phase space and size of the wave packet of a Bloch electron has non-zero lower bound due to which it rotates about its center of mass position, which give rise to OMM.
The total local current in the presence of finite intrinsic OMM is given by\cite{niu2} 
\begin{equation}{\label{j}}
\centering
\begin{aligned}
{\mathbf{j}}^{\textnormal{local}} & = -e\sum_{\lambda=\pm1}\int[d\mathbf{k}]{D}_{\mathbf{k}}^{\lambda}({\dot{\mathbf{r}}}^{\lambda})\tilde{f}_{{\mathbf{r}},{\mathbf{k}}}^{\lambda}\\
& + \boldsymbol{\nabla}_{\mathbf{r}}\times \sum_{\lambda=\pm1}\int[d\mathbf{k}]{D}_{\mathbf{k}}^{\lambda}({\mathbf{m}_{{\mathbf{k}}}^{\lambda}}){\tilde{f}_{\textnormal{eq}}^{\lambda}}.
\end{aligned}
\end{equation}
The results for the electrical transport and the thermoelectric transport of this system are presented in the subsequent sections for the cases ${E_F}$ $>$ 0 and ${E_F}$ $<$ 0 separately. For ${E_F}$ $>$ 0, both the bands are included in the summation  over $\lambda$ whereas for ${E_F}$ $<$ 0, only the band corresponding to the $\lambda$ = $-$1 contributes which includes the summation over the branch index ($\eta$). For numerical calculation, we consider the  following parameters: effective mass of an electron $m^{*}$ = 0.1${m_e}$, ${m_e}$ is the free electron mass, $\alpha$ = $10^{-10}$ eV-m, ${E_F}$ = 18.6 meV (for ${E_F} >$ 0) and ${E_F}$ = $-$6 meV (for ${E_F} <$ 0).
\section{Electrical transport}\label{IV}
\begin{center}
\textbf{A. Formalism}
\end{center}
In a spatially uniform system, only the first term of Eq. (\ref{j}) survives which is same as Eq. (\ref{j_electric}). Substituting Eqs. (\ref{EOM1}) and (\ref{NDF}) in Eq. (\ref{j_electric}), we obtained the Berry curvature and OMM dependent electric conductivity to the first order of the electric field as:
\begin{equation}{\label{sigma_ij}}
\begin{centering}
\begin{aligned}
\sigma_{ij} & = -\frac{e^{2}}{\hbar}\sum_{\lambda=\pm1}\int[d\mathbf{k}]\epsilon_{ijl}(\Omega_{{\mathbf{k}}}^{{\lambda},l})\tilde{f}_{\textnormal{eq}}^{\lambda}\\
&+e^{2}\tau\sum_{\lambda=\pm1}\int[d\mathbf{k}]{({D}_{\mathbf{k}}^{\lambda})}^{-1}\bigg({\tilde{v}}_{i}^{\lambda}+\frac{e}{\hbar}{B_{i}}
({\tilde{\mathbf{v}}_{\mathbf{k}}^{\lambda}}\cdot \bm{\Omega_\mathbf{k}^{\lambda}})\bigg)\\
& \bigg({\tilde{v}}_{j}^{\lambda}+\frac{e}{\hbar}{B_{j}}
({\tilde{\mathbf{v}}_{\mathbf{k}}^{\lambda}}\cdot \bm{\Omega_\mathbf{k}^{\lambda}})\bigg)\bigg(-\frac{{\partial{\tilde{f}_{\textnormal{eq}}^{\lambda}}}}{\partial{{\tilde{E}}_{\mathbf{k}}^{\lambda}}} \bigg),
\end{aligned}
\end{centering}
\end{equation}
where $\epsilon_{ijl}$ is the Levi-Civita tensor and $\Omega_{{\mathbf{k}}}^{{\lambda},l}$ denotes the $l$-component of Berry curvature. The first term refers to the intrinsic anomalous Hall effect which is independent of any scattering process (relaxation time). It arises due to presence of the Berry curvature without any magnetic field.

We can express the magnetoconductivity in terms of power of magnetic field by separating various $B$-contributions. The term $({D}_{\mathbf{k}}^{\lambda})^{-1}$ has been expanded upto quadratic order in $B$. The inverse expansion of $D$ converges when $\left(\frac{e{B}}{2 \hbar ({{k}_{F}^{\lambda}})^2}\right) < 1$. In our analysis, we have considered the carrier density $n_{e}= 10^{24}$ m$^{-3}$ and the maximum value of ${B} \sim$  5 T which gives the value of $\left(\frac{e{B}}{2 \hbar ({{k}_{F}^{\lambda}})^2}\right)\sim 0.25$ and 0.44 for $E_F>0$ and $E_F < 0$ respectively. So, the inverse expansion of $D$ in terms of $B$ holds good. The distribution function ($\tilde{f}_{\textnormal{eq}}^{\lambda}$) contains $B$-dependence through modified energy
 (${\tilde{E}}_{\mathbf{k}}^{\lambda}$), hence the Fermi function is expanded in terms of $B$ as
  $\tilde{f}_{\textnormal{eq}}^{\lambda}={f_{\textnormal{eq}}^{\lambda}}+( {\mathbf{m}}_\mathbf{k}^{\lambda}\cdot
  \mathbf{B})(\partial{f_{\textnormal{eq}}^{\lambda}}/\partial{E_{\mathbf{k}}^{\lambda}})$ with ${f_{\textnormal{eq}}^{\lambda}}$ is a distribution function when $B$ = 0\cite{moore}. We keep only first term of expansion in our calculations.

Expanding ${\sigma}_{ij}$ in Eq. ({\ref{sigma_ij}}) as:
\begin{equation}
{\sigma}_{ij}={\sigma}_{ij}^{\textnormal{AHE}}+{\sigma}_{ij}^{(0)}+ {\sigma}_{ij}^{(1)}+{\sigma}_{ij}^{(2)},
\end{equation}
where superscript indicates the order of magnetic field. The anomalous Hall conductivity is zero in our system.\\
The diagonal component of conductivity without magnetic field (known as Drude conductivity) is given by
\begin{equation}{\label{sigma_ij0}}
{\sigma}_{ij}^{(0)}= e^{2}\tau\sum_{\lambda=\pm1}\int[d\mathbf{k}] {{v}_{i}^{\lambda}}{{v}_{j}^{\lambda}}\left(-\frac{\partial{f_{\textnormal{eq}}^{\lambda}}}{\partial{E_{\mathbf{k}}^{\lambda}}}\right).
\end{equation}
The magnetoconductivity linear in magnetic field is given as:
\begin{equation}{\label{sigma_ij1}}
\centering
\begin{aligned}
{\sigma}_{ij}^{(1)} & =e^{2}\tau\sum_{\lambda=\pm1}\int[d\mathbf{k}] \bigg[({v_i^{\lambda}}{{v}_{j}^{m,{\lambda}}}+{v_j^{\lambda}}{{v}_{i}^{m,{\lambda}}})-\frac{e }{\hbar}( \mathbf{B} \cdot \bm{\Omega}_\mathbf{k}^{\lambda})\\&{v_i^{\lambda}}{v_j^{\lambda}}
+\frac{e}{\hbar}({v_i^{\lambda}}{B_j}+{v_j^{\lambda}}{B_i})(\mathbf{v}_\mathbf{k}^{\lambda}\cdot  \bm{\Omega}_\mathbf{k}^{\lambda})\bigg]
\left(-\frac{\partial{f_{\textnormal{eq}}^{\lambda}}}{\partial{E_{\mathbf{k}}^{\lambda}}}\right).
\end{aligned}
\end{equation}
The quadratic contribution of magnetic field to the conductivity as:
\begin{widetext}
\begin{equation}{\label{sigma_ij2}}
\begin{centering}
\begin{aligned}
{\sigma}_{ij}^{(2)} & = e^{2}\tau\sum_{\lambda=\pm1}\int[d\mathbf{k}] \Bigg[\frac{e^2 }{\hbar^2}{( \mathbf{B} \cdot \bm{\Omega}_\mathbf{k}^{\lambda})}^{2}{v_i^{\lambda}}{v_j^{\lambda}}-\frac{e }{\hbar}( \mathbf{B} \cdot \bm{\Omega}_\mathbf{k}^{\lambda})\bigg\{({v_i^{\lambda}}{{v}_{j}^{m,{\lambda}}}+{v_j^{\lambda}}{{v}_{i}^{m,{\lambda}}})+\frac{e }{\hbar} ({v_i^{\lambda}}{B_j}+{v_j^{\lambda}}{B_i})(\bf{v}_\mathbf{k}^{\lambda}\cdot  \bm{\Omega}_\mathbf{k}^{\lambda})\bigg\}\\
&+\bigg\{\frac{e}{\hbar}({v_i^{\lambda}}{B_j}+{v_j^{\lambda}}{B_i})({\bf{v}}_\mathbf{k}^{m,{\lambda}}\cdot  \bm{\Omega}_\mathbf{k}^{\lambda})+{\frac{e}{\hbar}({v}_{i}^{m,{\lambda}}}{B_j}+{v}_{j}^{m,{\lambda}}{B_i})({\bf{v}}_\mathbf{k}^{\lambda}\cdot  \bm{\Omega}_\mathbf{k}^{\lambda})+{v}_{i}^{m,{\lambda}}{v}_{j}^{m,{\lambda}}+\frac{e^2 }{\hbar^2}{( \mathbf{v}_\mathbf{k}^{\lambda} \cdot \bm{\Omega}_\mathbf{k}^{\lambda})}^{2}{B_i}{B_j}\bigg\}\Bigg]\left(-\frac{\partial{f_{\textnormal{eq}}^{\lambda}}}{\partial{E_{\mathbf{k}}^{\lambda}}}\right).
\end{aligned}
\end{centering}
\end{equation}
\end{widetext}
It is to be noted that external field will induce a shift in the chemical potential, $\delta\mu  \sim  B^2$ for the fixed carrier density which will result in additional contribution of the order of $B^2$ in the conductivities\cite{shift}. We are working in the zero temperature  limit and therefore, while performing the calculations, a derivative of the Fermi-Dirac distribution function is substituted by the Dirac-delta function.

We consider the two orientations of magnetic field i.e., ${\mathbf{B}}\perp {\hat{\mathbf{z}}}$ and ${\mathbf{B}}\parallel {\hat{\mathbf{z}}}$. The analytical expressions of different elements of the conductivity matrix is calculated for both the cases.
\begin{center}
\textbf{B. Results}
\end{center}
\textbf{Case 1: ${\mathbf{B}}\perp {\hat{\mathbf{z}}}$}

\begin{figure}[htbp]
\includegraphics[trim={2cm 4cm 1.5cm  2cm},clip,width=9
cm]{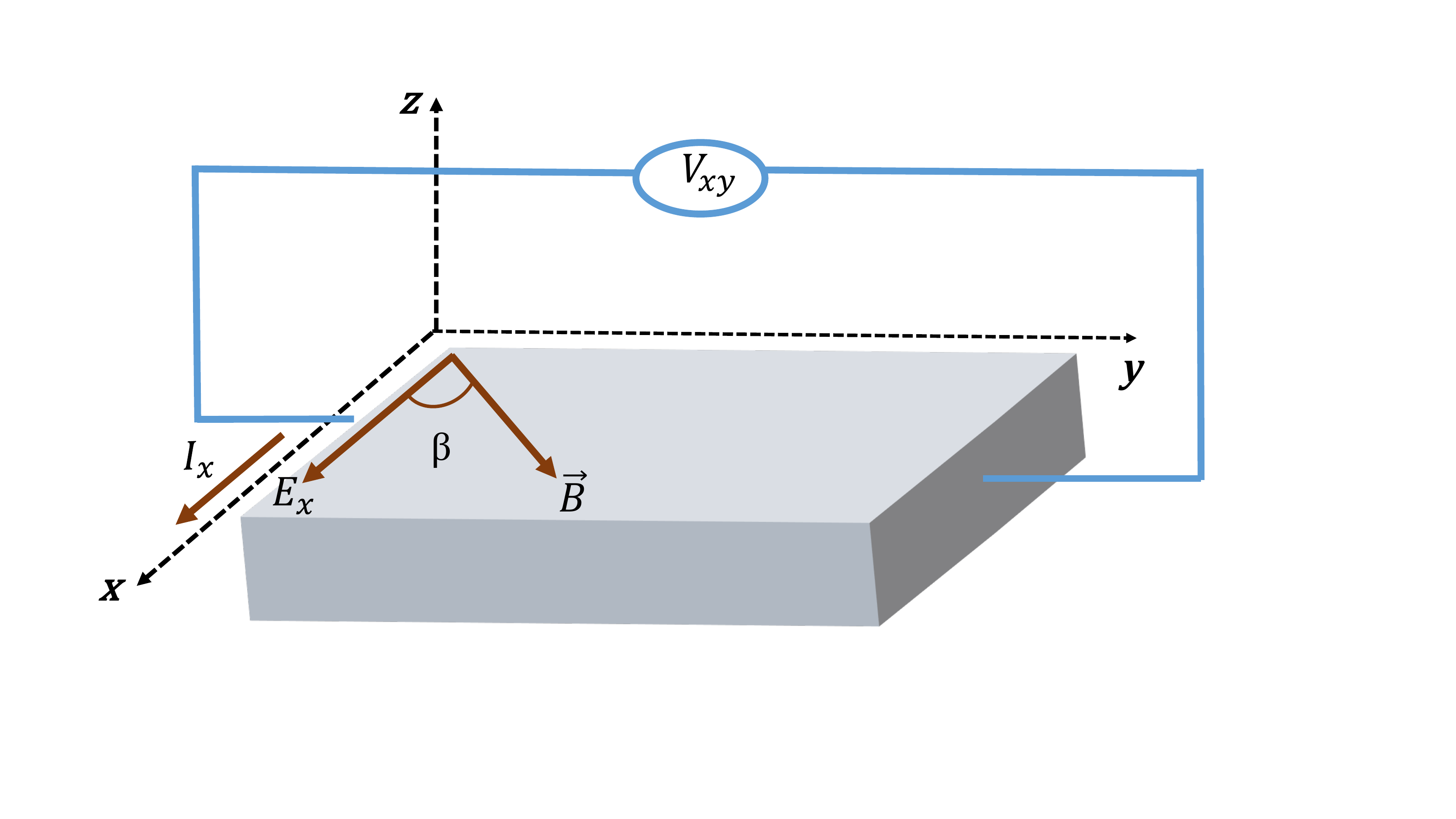}
\caption{Schematic illustration of the planar Hall effect geometry. The electric field $\mathbf{E}$ is applied along the $x$-axis and the magnetic field $\mathbf{B}$ is applied in the $x$-$y$ plane at an angle $\beta$ from the  $\mathbf{E}$. The planar Hall effect is measured as an in-plane induced voltage ($V_{xy}$) perpendicular to the direction of the external electric field.}
\label{01}
\end{figure}

We consider the externally applied electric field is along $x$-direction and the magnetic field is applied along the in-plane direction at an angle $\beta$ from the $x$ axis, i.e., ${\mathbf{B}}= (B\cos\beta, B\sin\beta, 0)$ as shown in Fig. \ref{01}. The electric field and magnetic field are perpendicular to each other at $\beta = \pi/2$.  
 The conductivity matrix equation takes the following form as
\begin{equation}{\label{matrix1}}
\left( \begin{matrix}
{J}_{x} \\
{J}_{y} \\
{J}_{z}
\end{matrix}\right)
= \left (\begin{matrix}
{\sigma_{xx}^{(0)}} + {\sigma_{xx}^{(2)}}&{\sigma_{xy}^{(2)}} &{0} \\
{\sigma_{yx}^{(2)}} & {\sigma_{yy}^{(0)}} + {\sigma_{yy}^{(2)}}  &{0}\\
{0} &{0}  & {\sigma_{zz}^{(0)}} + {\sigma_{zz}^{(2)}}
\end{matrix} \right)
\left( \begin{matrix}
{E}_{x} \\
{E}_{y} \\
{E}_{z}
\end{matrix}\right).
\end{equation}
The magnetic field independent conductivity (Drude conductivity) is obtained using Eq. (\ref{sigma_ij0}) as
\begin{equation}
\tilde{\sigma}_{xx}^{(0)}=\tilde{\sigma}_{D}= \left(\frac{2\sqrt{2}}{3 {\pi}^{2}}\bigg)\bigg[( \tilde{E}_{F}+2){\sqrt{\tilde{E}_{F}+1}}\right],
\end{equation}
and $\tilde{\sigma}_{yy}^{(0)}$=$\tilde{\sigma}_{zz}^{(0)}$=$\tilde{\sigma}_{xx}^{(0)}$. The analytical expression of quadratic in-plane diagonal component of the magnetoconductivity is explicitly given by using Eq. (\ref{sigma_ij2}) as
\begin{equation}{\label{longitudinal}}
\tilde{\sigma}_{xx}^{(2)} =\tilde{\sigma}_{\perp}^{(2)}+\Delta\tilde{\sigma}\cos^{2}\beta .
\end{equation}
\begin{figure}[htbp]
\hspace{0cm}\includegraphics[trim={2cm 2.5cm 1cm 1.5cm},clip,width=9cm]{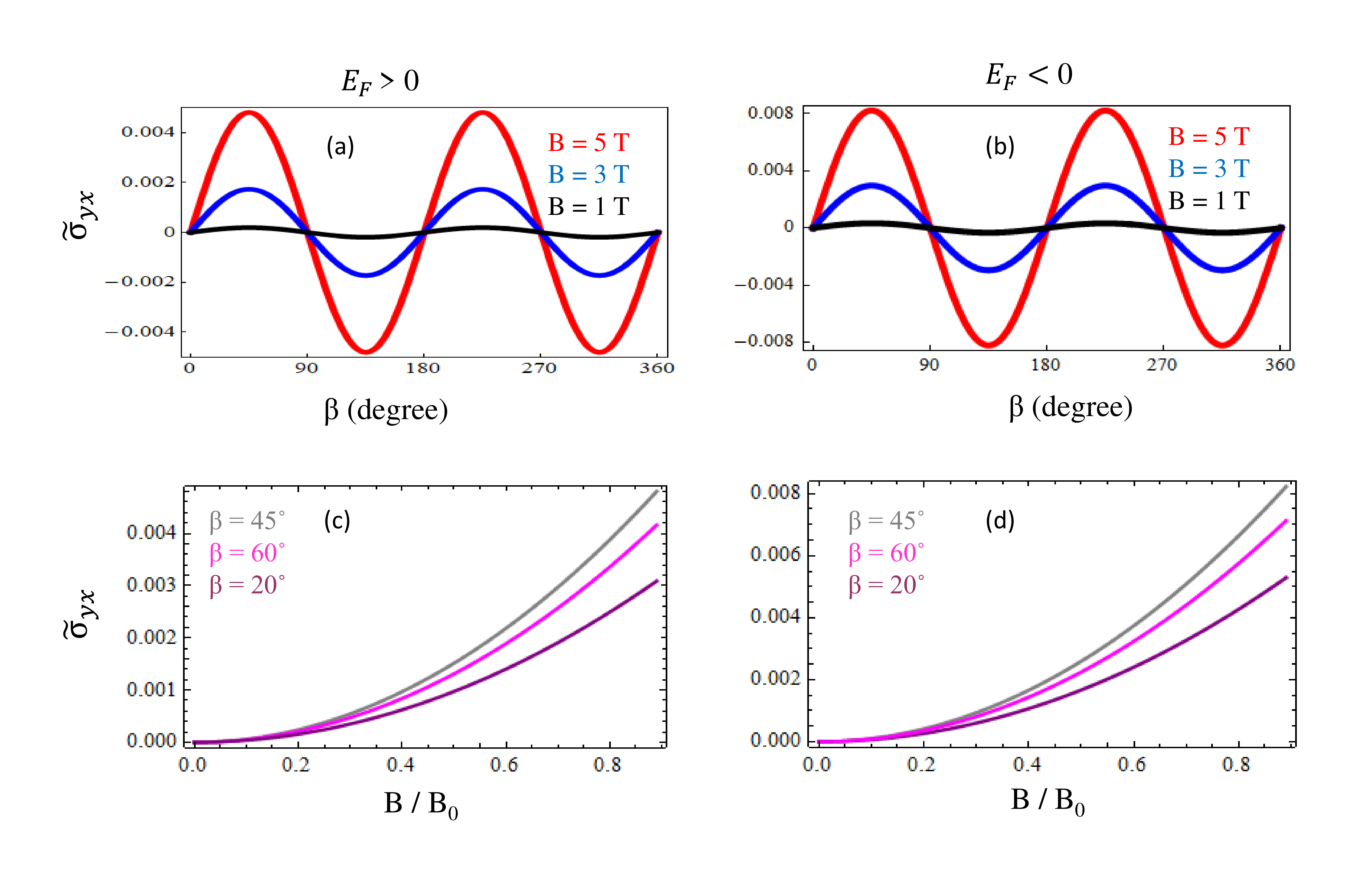}
\caption{Variation of the planar Hall conductivity as a function of angle between $\mathbf{E}$ and $\mathbf{B}$ in the planar geometry for $B$ = 5 T when (a) ${E_F} > 0$, (b) ${E_F} < 0$. The $B$-dependence of PHC at $\beta$ =$ \pi/4$ for (c) ${E_F}$ $>$ 0 and (d) ${E_F}$ $<$ 0.}
\label{PHC}
\end{figure}
Here, $\Delta\tilde{\sigma}=\tilde{\sigma}_{\parallel}-\tilde{\sigma}_{\perp}$ with $\tilde{\sigma}_{\parallel}= {\tilde{\sigma}_{xx}}(\beta=0)=\tilde{\sigma}_{D}+\tilde{\sigma}_{\parallel}^{(2)}$ and $\tilde{\sigma}_{\perp}={\tilde{\sigma}_{xx}}(\beta=\pi/2)= \tilde{\sigma}_{D}+\tilde{\sigma}_{\perp}^{(2)}$, where
\begin{equation}
\tilde{\sigma}_{\parallel}^{(2)}= \bigg(\frac{\sqrt{2}{\tilde{B}}^2}{120 {\pi}^{2}}\bigg)\Bigg[\frac{\big(8 \tilde{E}_{F}^{2}+27\tilde{E}_{F}+26\big)}{\tilde{E}_{F}^{2}\sqrt{\tilde{E}_{F}+1}} \Bigg],
\end{equation}
\begin{equation}
\tilde{\sigma}_{\perp}^{(2)}= \bigg(\frac{\sqrt{2}{\tilde{B}}^2}{120 {\pi}^{2}}\bigg)\Bigg[\frac{\big( \tilde{E}_{F}^{2}- \tilde{E}_{F}+2\big)}{\tilde{E}_{F}^{2}\sqrt{\tilde{E}_{F}+1}} \Bigg].
\end{equation}
Now, as expected,
\begin{equation}
\tilde{\sigma}_{yy}^{(2)}(\beta)=\tilde{\sigma}_{xx}^{(2)}(\pi/2-\beta).
\end{equation}
The zz-component of conductivity  quadratic in $B$ is given by
\begin{equation}
\tilde{\sigma}_{zz}^{(2)}= \bigg(\frac{\sqrt{2}{\tilde{B}}^2}{120 {\pi}^{2}}\bigg)\Bigg[\frac{\big( \tilde{E}_{F}^{2}- \tilde{E}_{F}+2\big)}{\tilde{E}_{F}^{2}\sqrt{\tilde{E}_{F}+1}} \Bigg].
\end{equation} 
We have defined $\tilde{{E_F}}={E_F}/{E_{\alpha}}$, $\tilde{B}={B}/B_{0}$ with $B_{0}=\left(m^{2}{\alpha}^{2}/{2e{\hbar}^{3}}\right)$ and $\tilde{\sigma}={\sigma}/\left(\frac{\tau{e^2}{E}_{\alpha}^{3/2}{m}^{1/2}}{{\hbar}^{3}}\right)$ as scaled Fermi energy, magnetic field and conductivity respectively. The linear dependence of conductivity in $B$ is zero.

\textbf{\textit{Planar Hall effect}}: We are interested in the planar Hall effect (PHE) observed in our case of planar geometry. The PHE\cite{nandy, burkov} appears in a configuration when the applied electric field, magnetic field and the induced voltage are co-planar such that the induced voltage is perpendicular to the electric field. The expression for planar Hall conductivity (PHC) is   
\begin{equation}
\tilde{\sigma}_{yx}= \bigg(\frac{\sqrt{2}{\tilde{B}}^2}{120 {\pi}^{2}}\bigg)\Bigg[\frac{\big(7 \tilde{E}_{F}^{2}+28\tilde{E}_{F}+24\big)}{\tilde{E}_{F}^{2}\sqrt{\tilde{E}_{F}+1}} \Bigg](\sin\beta \cos\beta).
\end{equation}
The planar Hall conductivity shows the $\sin\beta\cos\beta$ dependence, whereas longitudinal magnetoconductivity (LMC) follows the dependence of $\cos^{2}\beta$ as shown in Eq. (\ref{longitudinal}). The PHC attains the maximum value at an odd multiple of $\pi/4$. The angular dependence of  PHC is depicted in Fig. \ref{PHC}(a) and \ref{PHC}(b). The amplitude of PHC shows a quadratic dependence on $B$, i.e., $\Delta\sigma \propto B^{2}$ for any value of $ \beta $ except for $\beta = 0$ and $\beta = \pi/2$  which is shown in Fig. \ref{PHC}(c) and \ref{PHC}(d), whereas LMC follows the $B^2$ dependence except at $\beta = \pi/2$. We can also write PHC as:
\begin{equation}
\tilde{\sigma}_{yx}= \Delta\tilde{\sigma}\sin\beta \cos\beta.
\end{equation}
The PHC does not follow the anti-symmetry relation of regular Hall conductivity. Thus
$\tilde{\sigma}_{xy}=\tilde{\sigma}_{yx}$, as its origin is associated with the Berry curvature term and not to the Lorentz force. All the other off-diagonal components are zero. The expressions of magnetoconductivity for ${E_F} > 0$ and ${E_F} < 0$ holds the same form.

\textbf{\textit{Magnetoresistance}}: To study the effect of magnetic field, we calculate the magnetoresistance which is defined as 
\begin{equation}{\label{MR}}
\textnormal{MR}_{ii} = \frac{{\rho_{ii}(B)}-{\rho(0)}}{{\rho(0)}}, 
\end{equation}
\begin{figure}[htbp]
\hspace{0cm}\includegraphics[trim={1cm 4cm 0.5cm  2.5cm},clip,width=9cm]{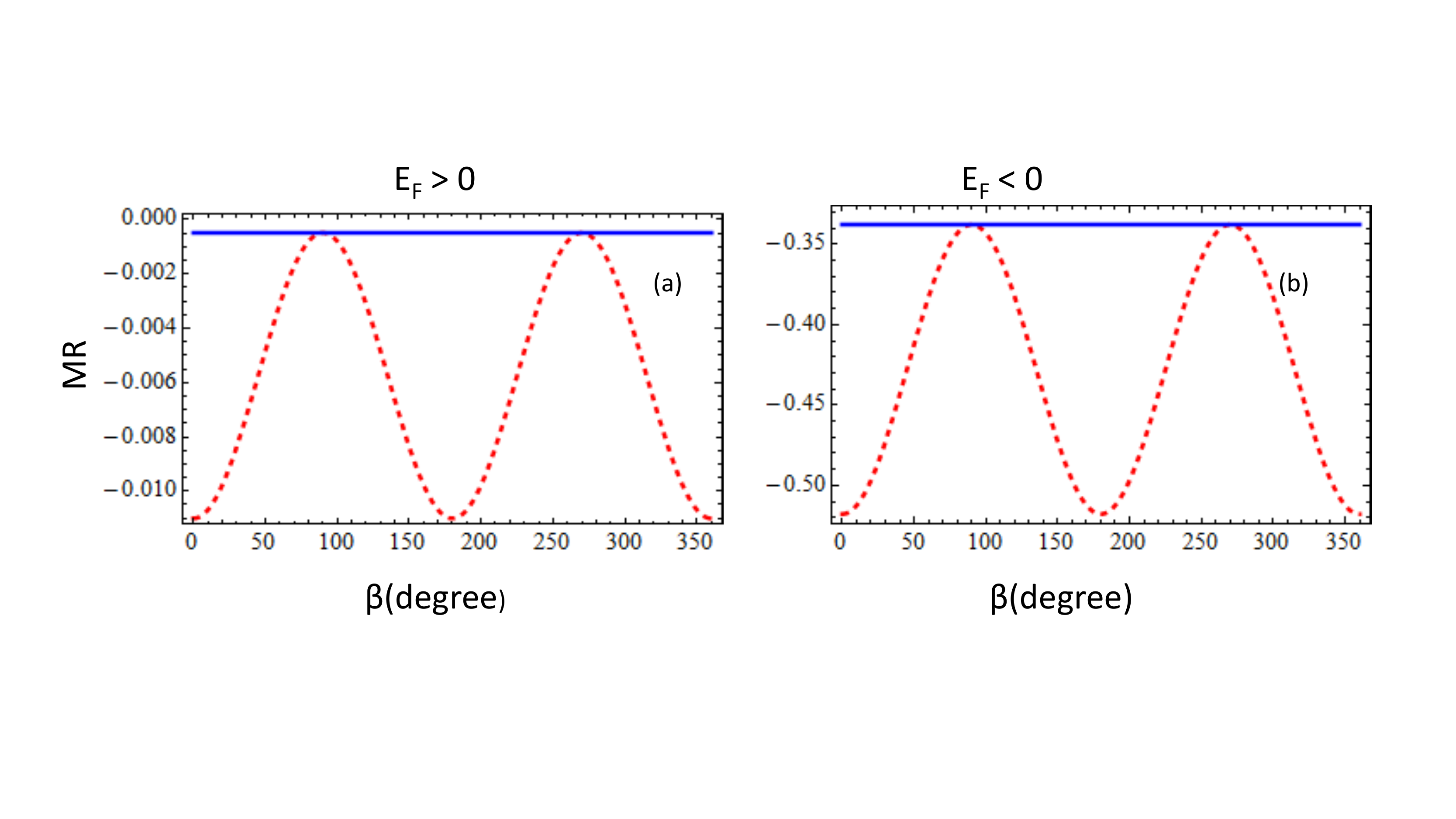}
\caption{The angular dependence of magnetoresistance of noncentrosymmetric metals at $B$ = 5 T for the planar geometry: (a) ${E_F} > 0$, (b) ${E_F} < 0$. In both (a) and (b), the red dashed curve shows the planar MR (MR$_{xx}(\beta)$) and the blue line depicts the out-of-plane MR (MR$_{zz})$.}
\label{21}
\end{figure}
where $i = x, y, z$. The Drude resistivity is defined as $\rho(0) = 1/{\sigma_{D}}$. The expression of planar MR ($\textnormal{MR}_{xx}(\beta)$) is given by Eq. (\ref{planar MR}). Here, the planar resistivity component $\rho_{xx}(\beta)$ is obtained by inverting the conductivity matrix. We observed that the magnetoconductivity increases monotonically with the magnetic field and follows the $B^2$ dependence due to the presence of Berry curvature and OMM leading to the decrease in magnetorestivity or negative MR. The planar MR [MR$_{xx}(\beta$)] follows the angular dependence of $\cos^{2}\beta$. The decrease in magnetoresistivity is maximum at $\beta= 0$ and $\pi$ and minimum at $\beta= \pi/2$. In Fig. \ref{21}, we have plotted the variation of planar MR with the angle between the electric field and magnetic field (red dashed curve). For ${E_F} > 0$, the negative MR resulting from the Berry curvature and OMM is about $-1\%$ at  ${E_F} = 18.4$ meV, $B$ = 5 T and $\beta= 0$, $\pi$ as shown in  Fig. \ref{21}(a). For ${E_F} < 0$, the MR reaches about $-50\%$ at ${E_F} = -6$ meV,  as shown in Fig. \ref{21}(b). Thus the effects of Berry curvature and OMM are considerably large for ${E_F} < 0$. This  difference in the nature of magnetoresistance is related to the magnitudes of the velocities, Berry curvature and OMM on the respective Fermi surfaces. However, it is to be noted that MR shows similar values below and above the BTP when effects of OMM are not included. Hence, OMM plays the major role in creating large difference in magnitudes of MR  below and above the BTP.

\begin{figure}[htbp]
\hspace{0cm}\includegraphics[trim={1cm 4cm 1cm  2.5cm},clip,width=9cm]{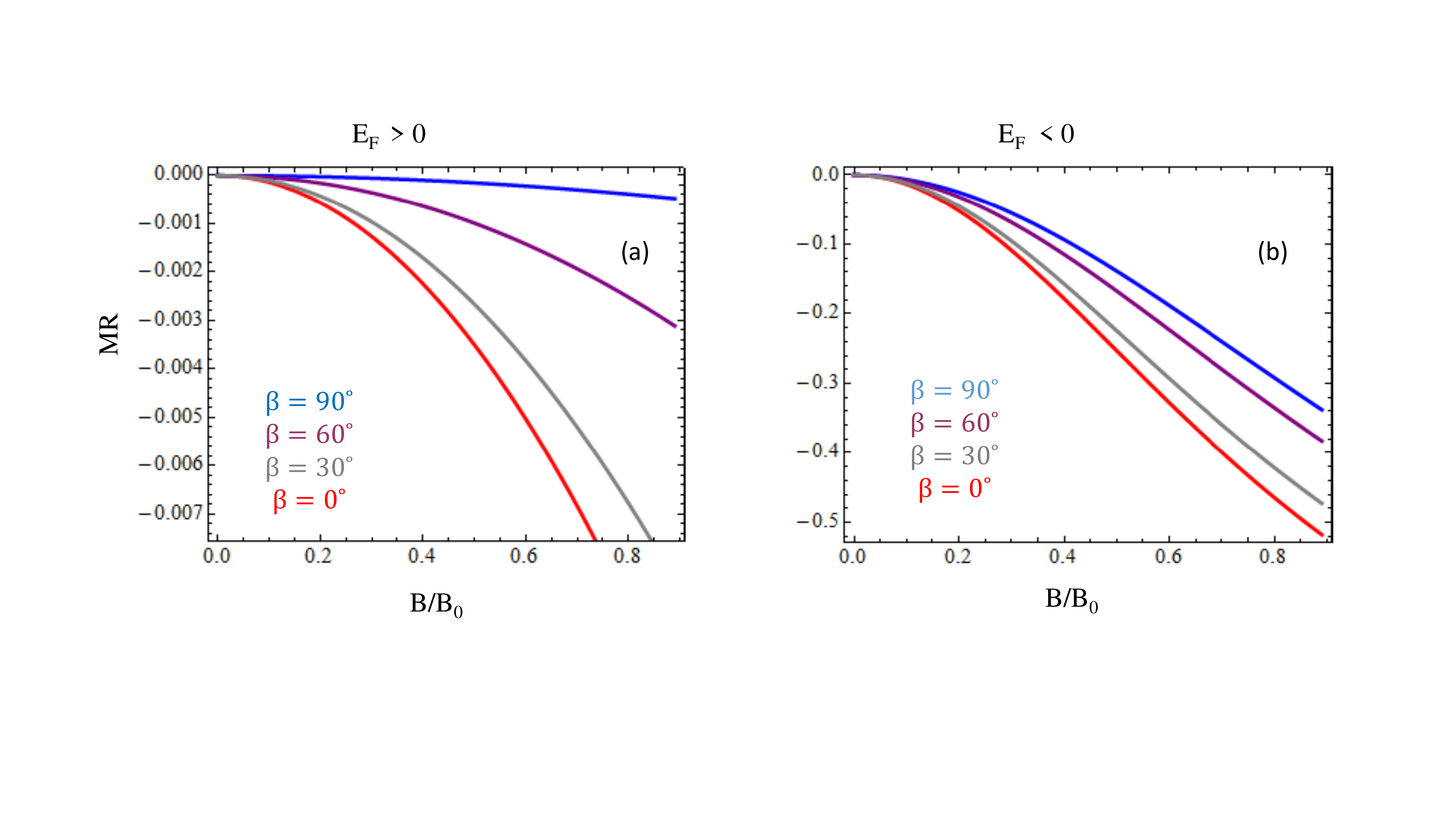}
\caption{The variation of magnetoresistance  for the planar geometry with the magnetic field for different angles between $\mathbf{E}$ and $\mathbf{B}$: (a) ${E_F} >$ 0, (b) ${E_F} <$ 0.}
\label{newplot}
\end{figure}
The out-of-plane MR is denoted by MR$_{zz}$ and its explicit expression is given by Eq. (\ref{outofplaneMR}). In Fig. \ref{21}, the blue solid lines show the variation of MR$_{zz}$  with the angle which appears to be constant as it is independent of angle ($\beta$).

The plots of variation of planar MR with the magnetic field for different angles between  $\mathbf{E}$ and $\mathbf{B}$ are shown in Fig. \ref{newplot}. For ${E_F} > 0$, when $B/{B_0} = 0.4$, the change in MR [MR$_{xx}(\beta=0$)$ - $ MR$_{xx}(\beta=\pi/2$)] resulting from the Berry curvature and OMM is about $-0.2\%$ as shown in  Fig. \ref{newplot}(a). When ${E_F} < 0$, the change in  MR reaches about $-9\%$ as shown in Fig. \ref{newplot}(b) at the same magnetic field.

\begin{figure}[htbp]
\hspace{0cm}\includegraphics[trim={1cm 1cm 0cm  0cm},clip,width=9cm]{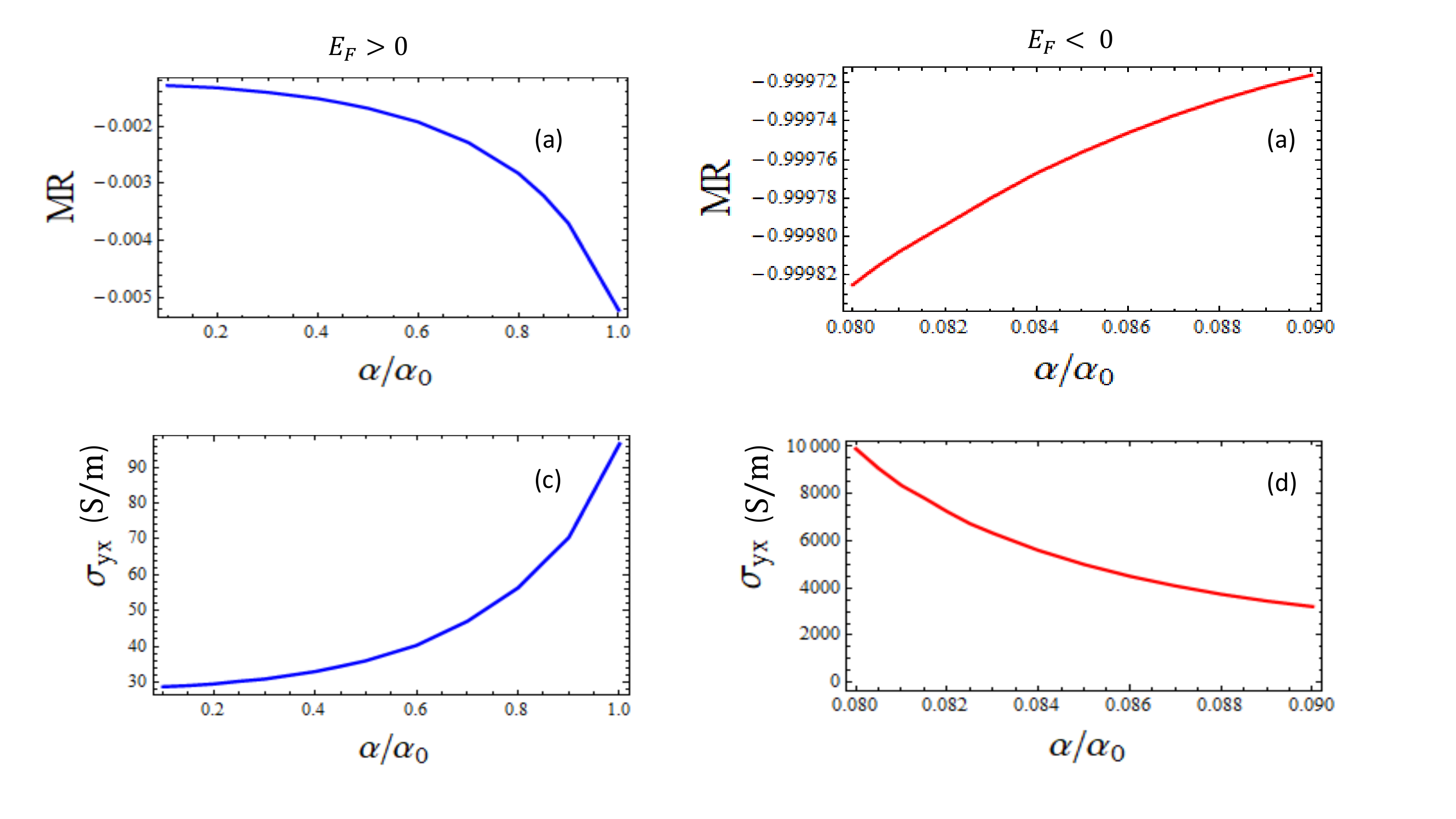}
\caption{The variation of magnetoresistance with $\alpha$ at $B$ = 5 T and $\alpha_{0}$ = $10^{-10}$ eV-m  for the cases: (a) ${E_F} >$ 0, (b) ${E_F} <$ 0. The ${\alpha}$-dependence of PHC for (c) ${E_F} >$ 0 and (d) ${E_F} <$ 0. }
\label{alpha}
\end{figure}

\textbf{\textit{Dependence on Rashba coupling parameter} ($\alpha$)}: The variation of magnetoresistance with $\alpha$ is plotted in Fig. \ref{alpha}(a) and \ref{alpha}(b). For ${E_F} > 0$, the absolute value of MR increases with $\alpha$ whereas for ${E_F} < 0$, it decreases with $\alpha$. The dependence of planar Hall conductivity on Rashba strength is depicted in Fig. \ref{alpha}(c) and \ref{alpha}(d).  When ${E_F} > 0$, the PHC increases with $\alpha$ whereas for ${E_F} < 0$, the PHC decreases with $\alpha$. 

The above results are obtained by neglecting the effect of Lorentz force. When $\mathbf{E}$ and $\mathbf{B}$ are parallel, the Lorentz force trivially vanishes and only the BC effects prevail. When there is a finite angle between $\mathbf{E}$ and $\mathbf{B}$, the Lorentz force leads to the additional corrections in the magnetoconductivities. However, it is known that Lorentz force does not contribute to the MR of single band systems with parabolic dispersion\cite{negmrniu}. Since the dispersion in this system is approximately parabolic for large $E_F$, the Lorentz force contribution in MR can be neglected.\\
\\
\textbf{Case 2: ${\mathbf{B}} \parallel {\hat{\mathbf{z}}}$}

For the magnetic field along $z$-direction, i.e., ${\mathbf{B}} =  B{\hat{\mathbf{z}}}$.
The conductivity matrix equation takes the following diagonal form: 
\begin{equation}\label{matrix3}
\left( \begin{matrix}
{J}_{x} \\
{J}_{y} \\
{J}_{z}
\end{matrix}\right)
= \left (\begin{matrix}
{\sigma_{xx}^{(0)}} + {\sigma_{xx}^{(2)}} &{0} &{0} \\
{0} & {\sigma_{yy}^{(0)}} + {\sigma_{yy}^{(2)}} &{0}\\
{0} &{0}  & {\sigma_{zz}^{(0)}} + {\sigma_{zz}^{(2)}}
\end{matrix} \right)
\left( \begin{matrix}
{E}_{x} \\
{E}_{y} \\
{E}_{z}
\end{matrix}\right).
\end{equation} 
The diagonal component of conductivity quadratic in $B$ is calculated to be
\begin{equation}
\tilde{\sigma}_{xx}^{(2)}=\tilde{\sigma}_{yy}^{{(2)}} =\bigg(\frac{\sqrt{2}{\tilde{B}}^2}{120 {\pi}^{2}}\bigg)\Bigg[\frac{\big( \tilde{E}_{F}^{2}- \tilde{E}_{F}+2\big)}{\tilde{E}_{F}^{2}{\sqrt{\tilde{E}_{F}+1}}} \Bigg],
\end{equation}
\begin{equation}
\tilde{\sigma}_{zz}^{(2)} =\bigg(\frac{\sqrt{2}{\tilde{B}}^2}{120 {\pi}^{2}}\bigg)\Bigg[\frac{\big( 8\tilde{E}_{F}^{2}+27\tilde{E}_{F}+26\big)}{\tilde{E}_{F}^{2}{\sqrt{\tilde{E}_{F}+1}}} \Bigg].
\end{equation} 
\begin{figure}[htbp]
\hspace{0cm}\includegraphics[trim={1cm 3cm 1cm  3cm},clip,width=9cm]{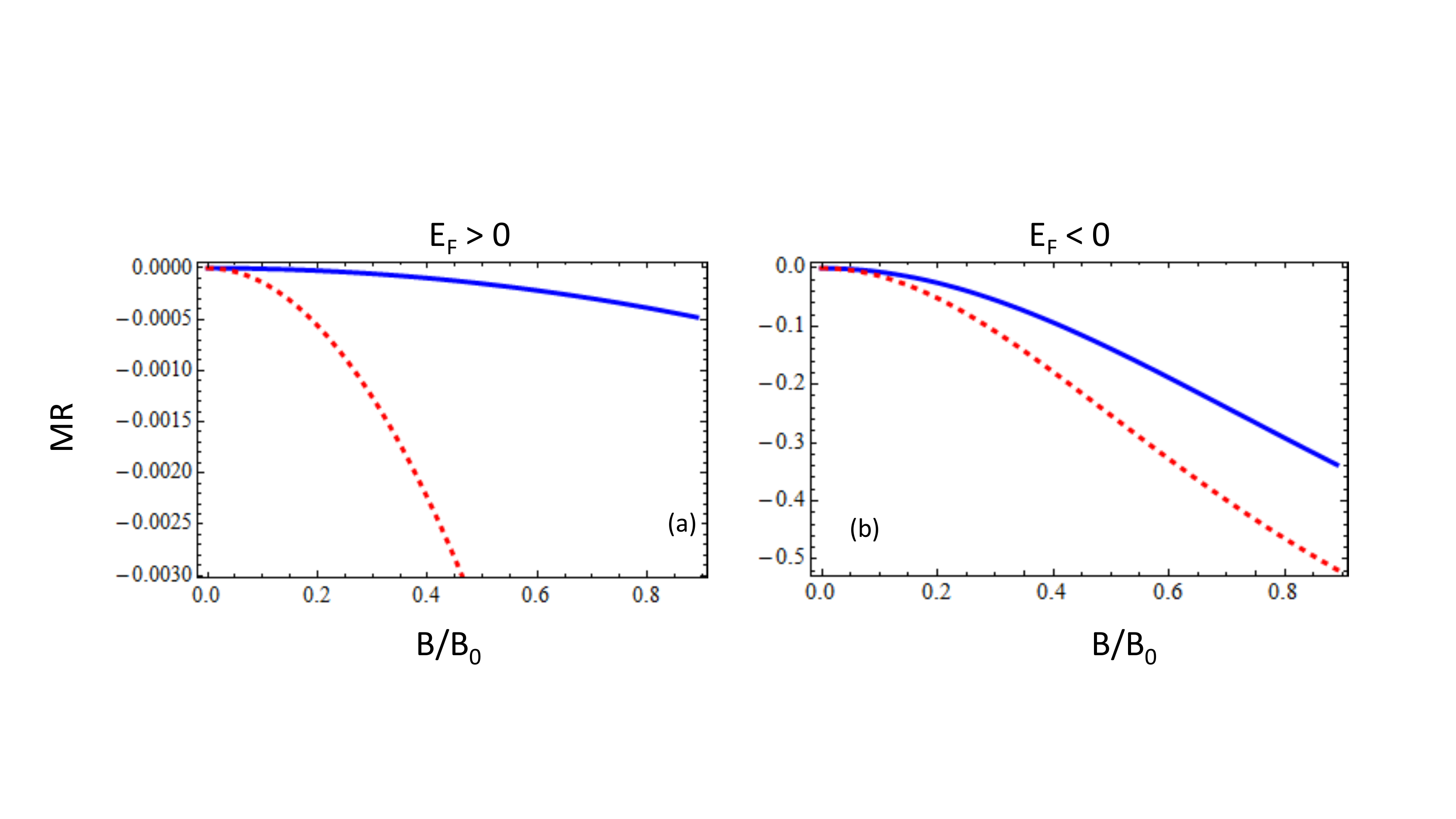}
\caption{Plot of  MR of noncentrosymmetric metals as a function of magnetic field for the ${\mathbf{B}}\parallel {\hat{\mathbf{z}}}$ case: (a) ${E_F} >$ 0, (b) ${E_F} <$ 0. The red dashed curve represents the longitudinal MR (MR$_{zz}$) and the blue curve depicts the perpendicular MR (MR$_{xx})$.}
\label{31}
\end{figure}
\textbf{\textit{Magnetoresistance}}: The longitudinal MR (MR$_{zz}$) is given by Eq. (\ref{MRzz}). And the expression of perpendicular MR (MR$_{xx}$) is obtained as Eq. (\ref{MRxx}). In Fig. \ref{31}(a), when ${E_F} > 0$, the negative longitudinal MR is about $-1\%$ and the perpendicular MR reaches around $-0.04\%$ at $B$ = 5 T. When ${E_F} < 0$, the longitudinal MR reaches about $-50\%$ and the perpendicular MR reaches about $-33\%$, thus it is clear that the Berry curvature and OMM effects on MR are considerably large  as shown in Fig. \ref{31}(b) at the same magnetic field.
\section{Thermoelectric transport}\label{V} 
\begin{center}
\textbf{A. Formalism}
\end{center}
The magnetization current in Eq. (\ref{j}) is not observable in conventional transport experiments, as it is localized current calculated from the self-rotation of the wave packet. Thus, it does not contribute to transport. Therefore, the transport current is defined as\cite{niu2,halperin}
\begin{equation}\label{jtrans}
{\mathbf{j}}^{\textnormal{trans}}={\mathbf{j}}^{\textnormal{local}}-\boldsymbol{\nabla}_{\mathbf{r}} \times \mathbf{M}(\mathbf{r}),
\end{equation}
where $\mathbf{M}(\mathbf{r})$ is the total orbital magnetization in real space. The transport current is related to the global motion of center of mass of the wave packet and contributes to the boundary current and is analogous to the free current in the classical electrodynamics\cite{jackson}.

The equilibrium magnetization density upto first order in magnetic field at finite temperatures is defined as\cite{niu2} 
\begin{equation}
F=-\frac{1}{\beta}\sum_{\lambda=\pm1}\int[d\mathbf{k}]{D_{\mathbf{k}}^{\lambda}} \ln[1+e^{-\beta({\tilde{E}}_{\mathbf{k}}^{\lambda}-\mu)}].
\end{equation}
The magnetization for the given chemical potential and temperature is given by $\mathbf{M}=-\left(\partial{F}/\partial{{\mathbf{B}}}\right)_{\mu,T} $ and
\begin{equation}\label{magnetization}
\begin{centering}
\begin{aligned}
\mathbf{M}(\mathbf{r}) & = \sum_{\lambda=\pm1} \int[d\mathbf{k}]{D_{\mathbf{k}}^{\lambda}}({\mathbf{m}_{{\mathbf{k}}}^{\lambda}})\tilde{f}_{\textnormal{eq}}^{\lambda}\\ 
& +\frac{1}{\beta}\sum_{\lambda=\pm1}\int[d\mathbf{k}]\bigg(\frac{e\bm{\Omega_\mathbf{k}^{\lambda}}}{\hbar}\bigg)\ln[1+e^{-\beta(\tilde{{E}}_{\mathbf{k}}^{\lambda}-\mu)}].
\end{aligned}
\end{centering}
\end{equation}
This is the general expression for equilibrium orbital magnetization density valid at non-zero magnetic field and arbitrary temperatures. Using Eqs. (\ref{j}) and (\ref{magnetization}) in Eq. (\ref{jtrans}), the transport current is given by
\begin{equation}
\centering
\begin{aligned}
{\mathbf{j}}^{\textnormal{trans}} & =-e\sum_{\lambda=\pm1}\int[d\mathbf{k}]{D_{\mathbf{k}}^{\lambda}}(\dot{{\mathbf{r}}}^{\lambda})\tilde{f}_{{\mathbf{r}},{\mathbf{k}}}^{\lambda}\\
& - \boldsymbol{\nabla}_{\mathbf{r}} \times\frac{1}{\beta}\sum_{\lambda=\pm1}\int[d\mathbf{k}]\left(\frac{e\bm{\Omega_\mathbf{k}^{\lambda}}}{\hbar}\right)\ln[1+e^{-\beta(\tilde{{E}}_{\mathbf{k}}^{\lambda}-\mu)}].
\end{aligned}
\end{equation}
The first term represents the usual charge current including the non-equilibrium correction to the first order in gradient of temperature which leads to the result depending on relaxation process (depending on $\tau$). The second term is the Berry phase correction to the magnetization. It is also defined upto first order in the statistical forces but independent of relaxation time and hence an intrinsic property of the system.

The Berry curvature and OMM dependent thermoelectric conductivity matrix defined linear in $\nabla{T}$ is obtained as
\begin{equation}\label{alpha_ij}
\begin{aligned}
\alpha_{ij} & = \frac{e{k_B}}{\hbar}\sum_{\lambda=\pm1}\int[d\mathbf{k}]\epsilon_{ijl}\Omega_{{\mathbf{k}}}^{{\lambda},l}\xi_{\mathbf{k}}^{\lambda}-e\tau\sum_{\lambda=\pm1}\int[d\mathbf{k}]{(D_{\mathbf{k}}^{\lambda})}^{-1}\\
&\frac{({\tilde{E}}_{\mathbf{k}}^{\lambda}-\mu)}{T}\bigg({\tilde{v}}_{i}^{\lambda}+\frac{e}{\hbar}{B_{i}}({\tilde{\mathbf{v}}_{\mathbf{k}}^{\lambda}}\cdot \bm{\Omega}_\mathbf{k}^{\lambda})\bigg)\\
& \bigg(  {\tilde{v}}_{j}^{\lambda}+\frac{e}{\hbar}{B_{j}} ({\tilde{\mathbf{v}}_{\mathbf{k}}^{\lambda}}\cdot \bm{\Omega}_\mathbf{k}^{\lambda})\bigg)
\bigg(-\frac{\partial{\tilde{f}_{\textnormal{eq}}^{\lambda}}}{\partial{{\tilde{E}}_{\mathbf{k}}^{\lambda}}} \bigg),
\end{aligned}
\end{equation}
with $ \xi_{\mathbf{k}}^{\lambda}$ defined as $\xi_{\mathbf{k}}^{\lambda}=\beta({\tilde{E}}_{\mathbf{k}}^{\lambda}-\mu)\tilde{f}_{\textnormal{eq}}^{\lambda} + \ln[1+e^{-\beta(\tilde{{E}}_{\mathbf{k}}^{\lambda}-\mu)}]$. The first term in Eq. (\ref{alpha_ij}) describes the purely anomalous thermoelectric effect\cite{niu2,kamal2} in the absence of magnetic field.\\\\
\begin{figure}[htbp]
\includegraphics[trim={1cm 2cm 1cm  1cm},clip,width=9
cm]{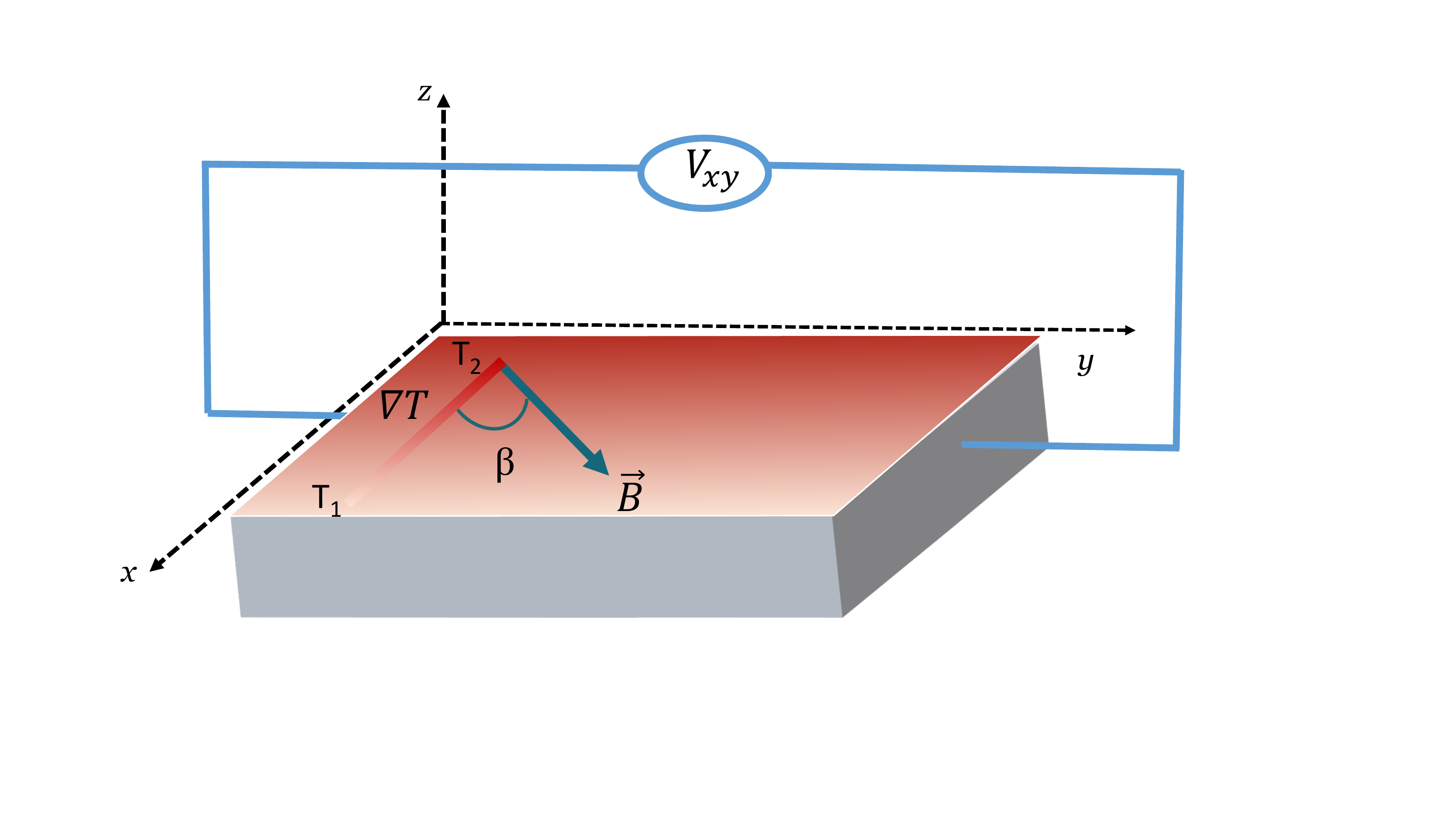}
\caption{Schematic illustration for the measurement of the planar Nernst coefficent in noncentrosymmetric metals. A temperature gradient ${dT}/{dx}$ produces a BC induced transverse electric field due to the co-planar component of the magnetic field.}
\label{te}
\end{figure} 
In the low-temperature limit $({k_{B}}T\ll {E_F})$, the thermoelectric conductivity can be expressed in terms of electrical conductivity using Mott relation\cite{ashcroft,niu2} given as
\begin{equation}\label{mottrelation}
\alpha_{ij}=\frac{-{\pi}^2{k}_{B}^{2}T}{3e}\bigg(\frac{\partial{\sigma_{ij}}}{\partial{{{E}}}}\bigg)_{E={E_F}}.
\end{equation}
\textbf{\textit{Thermopower}}: The thermopower is defined for an open circuit and therefore, we will keep electric current to be zero in Eq. (\ref{j_i}) and thus, the temperature gradient generates the electric field as
\begin{equation}\label{nu}
{E}_{i}=\nu_{ij}(\nabla_{j}T).
\end{equation}
The thermopower matrix can be evaluated using $\nu=\sigma^{-1}\alpha$. The diagonal components $(\nu_{ii})$ denotes the conventional thermopower or  Seebeck coefficient which describes the electric response along the direction of temperature gradient. The off-diagonal components  $(\nu_{ij})$ denotes the Nernst coefficients (NC). An anomalous (conventional) Nernst effect measures the thermoelectric voltage induced transverse to the temperature gradient in the presence of out-of-plane Berry curvature (magnetic field). The planar Nernst effect\cite{ANE3,kamal2} occurs in the configuration when the temperature gradient, magnetic field and the induced voltage are co-planar such that the voltage induced is transverse to the temperature gradient as depicted in Fig. \ref{te}. The planar component of $\mathbf{B}$ due to BC gives rise to PNE (also known as transverse thermopower).

\begin{center}
\textbf{B. Results}
\end{center}
\textbf{Case 1: ${\mathbf{B}}\perp {\hat{\mathbf{z}}}$}

The thermoelectric conductivity matrix takes the following form
\begin{equation}\label{matrixalpha}
\alpha
= \left (\begin{matrix}
{\alpha_{D}+{\alpha}_{xx}^{(2)}} &{\alpha_{xy}^{(2)}}&{0} \\
{\alpha_{yx}^{(2)}} & {\alpha_{D}+\alpha_{yy}^{(2)}}&{0}\\
{0}&{0}&{\alpha_{D}+{\alpha}_{zz}^{(2)}}
\end{matrix} \right).
\end{equation} 
The different elements of thermoelectric conductivity matrix can be obtained from the corresponding elements of conductivity matrix, i.e., Eq. (\ref{matrix1}) using Eq. (\ref{mottrelation}) as
\begin{equation}
\tilde{\alpha}_{D}=-\frac{2\sqrt{2}}{18}\left( \frac{3\tilde{E}_{F}+4}{\sqrt{(\tilde{E}_{F}+1)}}\right),
\end{equation}
\begin{widetext}
\begin{equation}
\tilde{\alpha}_{xx}^{(2)} = -\frac{\sqrt{2}{\tilde{B}}^{2}}{720}\Bigg[\Bigg( \frac{-\tilde{E}_{F}^{4}+3\tilde{E}_{F}^{3}-8\tilde{E}_{F}^{2}-8\tilde{E}_{F}}{{\tilde{E}_{F}^{4}}{(\tilde{{E_F}}+1)}^{3/2}}\Bigg)\sin^{2}\beta + \Bigg( \frac{-8\tilde{E}_{F}^{4}-81\tilde{E}_{F}^{3}-184\tilde{E}_{F}^{2}-104\tilde{E}_{F}}{\tilde{E}_{F}^{4}{(\tilde{E}_{F}+1)}^{3/2}}\Bigg)\cos^{2}\beta\Bigg],
\end{equation}
$$\tilde{\alpha}_{yy}^{(2)}(\beta)=\tilde{\alpha}_{xx}^{(2)}(\pi/2-\beta),$$ and 
\begin{equation}
\tilde{\alpha}_{zz}^{(2)}=-\frac{\sqrt{2}{\tilde{B}}^{2}}{720}\Bigg( \frac{-\tilde{E}_{F}^{4}+3\tilde{E}_{F}^{3}-8\tilde{E}_{F}^{2}-8\tilde{E}_{F}}{\tilde{E}_{F}^{4}{(\tilde{E}_{F}+1)}^{3/2}}\Bigg),
\end{equation} 
\begin{equation}
\tilde{\alpha}_{xy}^{(2)}=\tilde{\alpha}_{yx}^{(2)}=-\frac{\sqrt{2}{\tilde{B}}^{2}}{720}\Bigg( \frac{-7\tilde{E}_{F}^{4}-84\tilde{E}_{F}^{3}-176\tilde{E}_{F}^{2}-96\tilde{E}_{F}}{\tilde{E}_{F}^{4}{(\tilde{E}_{F}+1)}^{3/2}}\Bigg)\cos\beta\sin\beta.
\end{equation}
\end{widetext}
 The thermopower matrix is given as:
\begin{equation}
\nu
= \left (\begin{matrix}
{\nu_{D}+{\nu}_{xx}^{(2)}} &{\nu_{xy}^{(2)}}&{0} \\
{\nu_{yx}^{(2)}} & {\nu_{D}+\nu_{yy}^{(2)}}&{0}\\
{0}&{0}&{\nu_{D}+{\nu}_{zz}^{(2)}}
\end{matrix} \right),
\end{equation} 
where the various elements of thermopower matrix can be obtained using Eqs. (\ref{matrix1}) and (\ref{matrixalpha}) in Eq. (\ref{nu}) as 
\begin{equation}
\tilde{\nu}_{D}=\frac{\tilde{\alpha}_{D}}{\tilde{\sigma}_{D}} = -\frac{\pi^{2}}{6}\Bigg( \frac{3\tilde{E}_{F}+4}{(\tilde{E}_{F}+1)(\tilde{E}_{F}+2)}\Bigg).
\end{equation}
\begin{figure}[htbp]
\includegraphics[trim={0cm 0cm 0cm  0cm},clip,width=9cm]{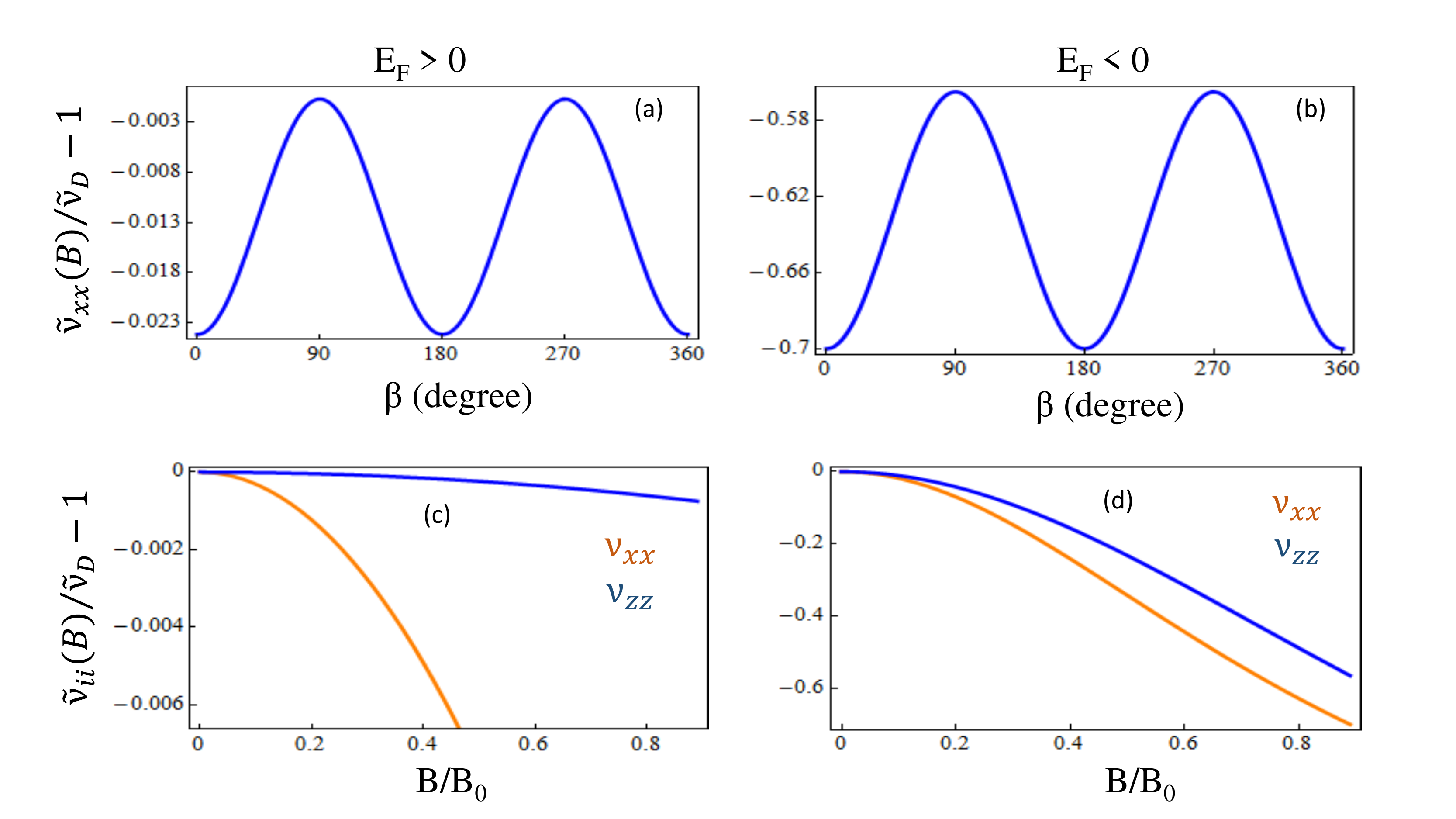}
\caption{The angular dependence of the planar SC for $B$ = 5 T in the  planar geometry when (a) ${E_F} >$ 0, (b) ${E_F} <$ 0. The dependence of the longitudinal SC (orange curve) and the out-of-plane SC (blue curve) with the magnetic field for (c) ${E_F}$ $>$ 0 and (d) ${E_F}$ $<$ 0.}
\label{51}
\end{figure}
We have defined  $\tilde{\alpha}$ = ${\alpha}/\left(\frac{e\tau{K}_{B}^{2}T{E}_{\alpha}^{1/2}{m}^{1/2}}{{\hbar}^{3}}\right)$ and $\tilde{\nu}$ = ${\nu}/\left(\frac{{K}_{B}^{2}T}{e{E}_{\alpha}}\right)$ as scaled thermoelectric conductivity and thermopower respectively.

The SC in the planar configuration ($\tilde{\nu}_{xx}(\beta)$) is given by Eq. (\ref{planarsc}) with $\tilde{\nu}_{yy}(\beta)=\tilde{\nu}_{xx}(\pi/2-\beta)$. The angular dependence of the planar SC is given by Fig. \ref{51}(a) and \ref{51}(b). The out-of-plane SC ($\nu_{zz}$) is given by Eq. (\ref{outofplanesc}). The longitudinal SC ($\tilde{\nu}_{xx}(\beta = 0)$) and the out-of-plane SC dependence on the magnetic field is depicted in Fig. \ref{51}(c) and \ref{51}(d). The magnetic field reduces the Seebeck coefficient in presence of Berry curvature, which results in negative Seebeck effect. 

The planar Nernst coefficient obtained for the planar configuration is given by Eq. (\ref{pnc}). The PNC shows the same angular dependence of $\cos\beta\sin\beta$ as in the case of planar Hall conductivity, therefore, it is finite in all directions except at $\beta=0$ and $\pi/2$ and attains the maximum at an odd multiple of $\pi/4$. Fig. \ref{NC}(a) and \ref{NC}(b) shows the dependence of PNC on the planar angle between $\mathbf{E}$ and $\mathbf{B}$ and the $B$-dependence of PNC is depicted in Fig. \ref{NC}(c) and \ref{NC}(d).\\
\begin{figure}[htbp]\label{NC}
\hspace{0cm}\includegraphics[trim={0.3cm 0cm 0cm  0cm},clip,width=9cm]{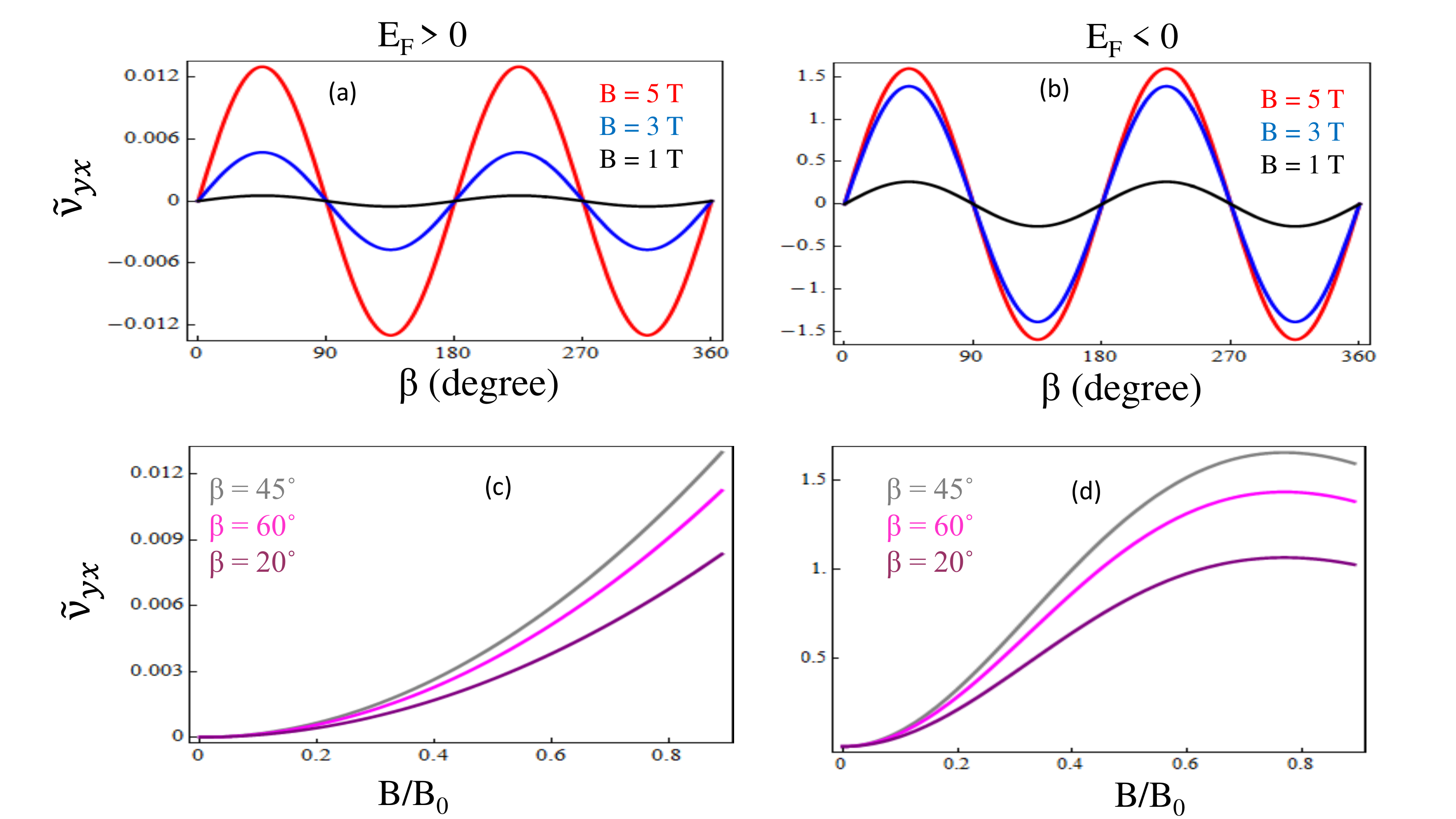}
\caption{The angular dependence of planar NC with the angle between the $\mathbf{E}$ and $\mathbf{B}$ for the planar geometry: (a) ${E_F} >$ 0, (b) ${E_F} <$ 0. The plots (c) and (d) shows the variation of the planar NC with the magnetic field for ${E_F} >$0 and ${E_F} <$ 0 respectively.}
\label{NC}
\end{figure}

\textbf{Case 2: ${\mathbf{B}}\parallel {\hat{\mathbf{z}}}$}

The thermoelectric conductivity matrix takes the following diagonal form:
\begin{equation}\label{matrix4}
\alpha
= \left (\begin{matrix}
{\alpha_{D}+{\alpha}_{xx}^{(2)}} &{0}&{0} \\
{0} & {\alpha_{D}+\alpha_{yy}^{(2)}}&{0}\\
{0}&{0}&{\alpha_{D}+{\alpha}_{zz}^{(2)}}
\end{matrix} \right),
\end{equation} 
\begin{figure}[htbp]
\includegraphics[trim={1cm 3cm 0cm  2cm},clip,width=9cm]{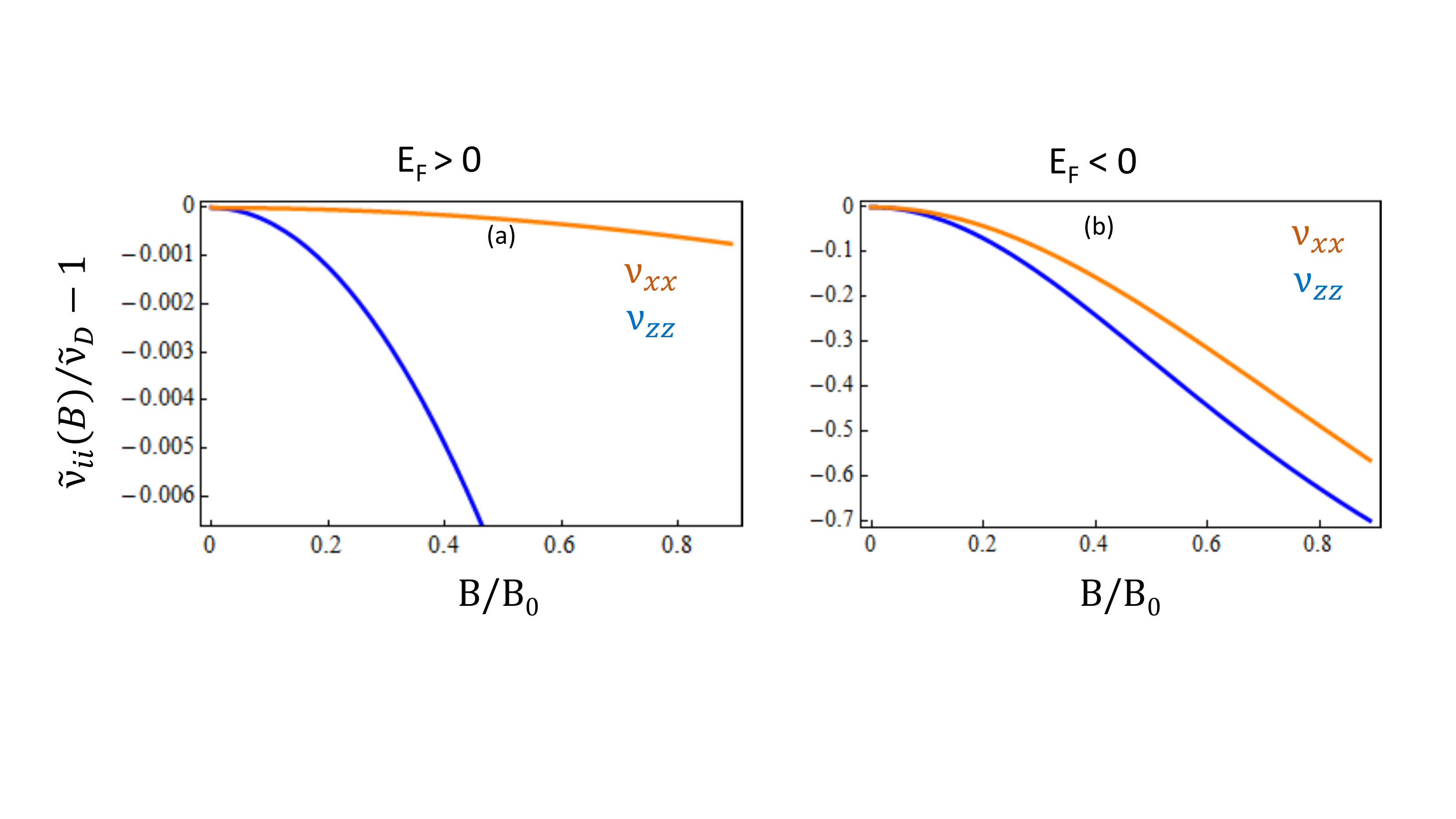}
\caption{Dependence of the SC as a function of magnetic field for the ${\mathbf{B}}\parallel {\hat{\mathbf{z}}}$ case: (a) ${E_F} >$ 0, (b) ${E_F} <$ 0. The blue curve represents the longitudinal SC ($\nu_{zz}$)  and the orange curve depicts the perpendicular SC ($\nu_{xx}$).}
\label{61}
\end{figure}
where
\begin{equation}
\tilde{\alpha}_{xx}^{(2)}=\tilde{\alpha}_{yy}^{(2)}=-\frac{\sqrt{2}{\tilde{B}}^{2}}{720}\Bigg[ \frac{-\tilde{E}_{F}^{4}+3\tilde{E}_{F}^{3}-8\tilde{E}_{F}^{2}-8\tilde{E}_{F}}{\tilde{E}_{F}^{4}{(\tilde{E}_{F}+1)}^{3/2}}\Bigg].
\end{equation} 
\begin{equation}
\tilde{\alpha}_{zz}^{(2)}=-\frac{\sqrt{2}{\tilde{B}}^{2}}{720}\Bigg[ \frac{-8\tilde{E}_{F}^{4}-81\tilde{E}_{F}^{3}-184\tilde{E}_{F}^{2}-104\tilde{E}_{F}}{\tilde{E}_{F}^{4}{(\tilde{{E_F}}+1)}^{3/2}}\Bigg].
\end{equation}

The thermopower matrix is obtained using Eqs. (\ref{matrix3}) and (\ref{matrix4}) in Eq. (\ref{nu}) as:
\begin{equation}
\nu
= \left (\begin{matrix}
{\nu_{D}+{\nu}_{xx}^{(2)}} &{0}&{0} \\
{0} & {\nu_{D}+\nu_{yy}^{(2)}}&{0}\\
{0}&{0}&{\nu_{D}+{\nu}_{zz}^{(2)}}
\end{matrix} \right),
\end{equation} 

The thermopower matrix has no off-diagonal components in this parallel configuration, therefore no Nernst coefficients. The diagonal components, i.e., the longitudinal SC $(\tilde{\nu}_{zz})$ and the perpendicular SC $(\tilde{\nu}_{xx})$  obtained are given by Eqs. (\ref{longitudinalsc}) and (\ref{perpendicularsc}) respectively (figure \ref{61}).
\section{Conclusion}\label{VI}
In this work, we have studied the magnetoelectric and  magnetothermal transport phenomena in noncentrosymmetric metals using semiclassical Boltzmann transport formalism by incorporating the effects of BC and OMM. The OMM enters the energy-dispersion relation in the form of Zeeman-like coupling term which modifies the velocity of the Bloch electrons. We have worked out the magnetoconductivity matrix for two orientations of magnetic field with respect to the $z$-axis (representing the unit normal to the current-voltage plane) -- ${\mathbf{B}}\perp {\hat{\mathbf{z}}}$ and ${\mathbf{B}}\parallel {\hat{\mathbf{z}}}$. 

We find that the magnetoconductivity increases monotonically with the magnetic field and follows the $B^2$ dependence due to the presence of Berry curvature and OMM resulting in negative MR for both the orientations. For the case of ${\mathbf{B}}\perp {\hat{\mathbf{z}}}$, the planar Hall effect is observed in the system. It is distinct from the Berry curvature induced anomalous Hall effect and the Lorentz force mediated Hall effect as the transverse conductivities corresponding to these two effects are antisymmetric in spatial indices.

The PHC and MR show the usual angular dependence of $\sin\beta\cos\beta$ and $\cos^{2}\beta$ respectively as shown in previous works\cite{nandy,kamal2}. For ${E_F}$ close to $-{E_\alpha}$, we get giant negative magnetoresistance with maximum value of about $-50\%$ (for ${\mathbf{E}} \parallel {\mathbf{B}}$) and minimum value of $-33\%$ (for ${\mathbf{E}} \perp {\mathbf{B}}$) for typical values of Rashba strength and when  ${E_F} <$ 0, the maximum MR is about $-1\%$. This  difference in the nature of magnetoresistance below and above the BTP is related to the magnitudes of the velocities, Berry curvature and OMM on the respective Fermi surfaces, where OMM plays the key role. For the case ${\mathbf{B}} \parallel {\hat{\mathbf{z}}}$, the conductivity matrix takes the diagonal form with quadratic-$B$ dependence components implying the absence of any Hall response. The absolute magnetoresistance and planar Hall conductivity show a decreasing (increasing) trend with Rashba coupling parameter for Fermi energy lesser
(greater) than zero.  

The thermopower matrix is obtained from the conductivity matrix using Mott relation. For the case of ${\mathbf{B}}\perp {\hat{\mathbf{z}}}$, we note that the PNE and negative Seebeck effect have same angular dependence as that of PHC and MR respectively. When ${\mathbf{B}}\parallel {\hat{\mathbf{z}}}$, the conductivity matrix takes the diagonal form with no Nernst coefficient.  

\begin{center}
	{\bf ACKNOWLEDGEMENTS}
\end{center}
We would like to thank Sonu Verma and Pooja Kesarwani for
useful discussions.

\begin{widetext}
\appendix{}

\section{Electrical transport}
Here, we present the final expressions of magnetoresistance for the different orientations of magnetic field.\\
For the case of ${\mathbf{B}}\perp {\hat{\mathbf{z}}}$ (planar geometry), the planar MR is calculated as
\begin{equation}\label{planar MR}
\centering
\begin{aligned}
\textnormal{MR} & = -1+\bigg[40  \tilde{E}_{F}^2 \left(2+3 \tilde{E}_{F}+ \tilde{E}_{F}^2\right) \bigg\{160  \tilde{E}_{F}^2 \left(2+3  \tilde{E}_{F}+ \tilde{E}_{F}^2\right)+\tilde{B}^2 \left(28+26  \tilde{E}_{F}+9  \tilde{E}_{F}^2\right) \\
&-\tilde{B}^2 \left(24+28  \tilde{E}_{F}+7  \tilde{E}_{F}^2\right)\cos{2 \beta}\bigg\}\bigg]\bigg/\bigg[6400  \tilde{E}_{F}^4 \left(2+3 \tilde{E}_{F}+ \tilde{E}_{F}^2\right)^2+\tilde{B}^4 \big(52+28  \tilde{E}_{F}\\
&+15  \tilde{E}_{F}^2+19  \tilde{E}_{F}^3+8  \tilde{E}_{F}^4\big)
+80 \tilde{B}^2  \tilde{E}_{F}^2 \left(56+136  \tilde{E}_{F} + 124  \tilde{E}_{F}^2+53  \tilde{E}_{F}^3+9  \tilde{E}_{F}^4\right)\bigg].
\end{aligned}
\end{equation}
The out-of-plane MR is obtained as
\begin{equation}\label{outofplaneMR}
\textnormal{MR}_{zz}=-1+\frac{80  \tilde{E}_{F}^2 (1+ \tilde{E}_{F}) (2+ \tilde{E}_{F})}{\tilde{B}^2 \left(2- \tilde{E}_{F}+ \tilde{E}_{F}^2\right)+80  \tilde{E}_{F}^2 \left(2+3  \tilde{E}_{F}+ \tilde{E}_{F}^2\right)}.
\end{equation}
For the other case of ${\mathbf{B}}\parallel {\hat{\mathbf{z}}}$, we calculate the longitudinal MR to be
\begin{equation}\label{MRzz}
\textnormal{MR}_{zz}=-1+\frac{80\tilde{E}_{F}^2 \left(2+3 \tilde{E}_{F}+ \tilde{E}_{F}^2\right)}{80  \tilde{E}_{F}^2 \left(2+3  \tilde{E}_{F}+ \tilde{E}_{F}^2\right)+{\tilde{B}}^2 \left(26+27  \tilde{E}_{F}+8  \tilde{E}_{F}^2\right)}.
\end{equation}
The perpendicular MR is given by
\begin{equation}\label{MRxx}
\textnormal{MR}_{xx}=-1+\frac{80  \tilde{E}_{F}^2 \left(2+3  \tilde{E}_{F}+ \tilde{E}_{F}^2\right)}{{\tilde{B}}^2 \left(2- \tilde{E}_{F}+ \tilde{E}_{F}^2\right)+80  \tilde{E}_{F}^2 \left(2+3  \tilde{E}_{F}+ \tilde{E}_{F}^2\right)}.
\end{equation}

\section{Thermoelectric transport}
In this Appendix, we provide the results of thermoelectric transport coefficients for  the different configurations of the magnetic field.

For the case of ${\mathbf{B}}\perp \hat{\mathbf{z}}$ (planar geometry), the SC in the planar configuration is given by
\begin{equation}\label{planarsc}
\centering
\begin{aligned}
{\tilde{\nu}_{xx}} & = \Bigg[\pi ^2 \Bigg\{2 \Bigg(-3200 \tilde{E}_{F}^5 (1+ \tilde{E}_{F}^2 )\left(8+10  \tilde{E}_{F}+3 \tilde{E}_{F}^2\right)-40 {\tilde{B}}^2   \tilde{E}_{F}^2 (1+  \tilde{E}_{F})^2 \left(-112-80  \tilde{E}_{F}+9   \tilde{E}_{F}^3\right)\\
& +{\tilde{B}}^{4}\left(104+172   \tilde{E}_{F}+71  \tilde{E}_{F}^2+32   \tilde{E}_{F}^3+19   \tilde{E}_{F}^4+4
\tilde{E}_{F}^5\right)\Bigg) +5{\tilde{B}}^{2}  \tilde{E}_{F}(1+ \tilde{{E_F}})
\bigg[{\tilde{B}}^2 \left(-16+4  
\tilde{E}_F+7 \tilde{E}_F^2\right)\\
& +16  \tilde{E}_{F} \left(96+272  \tilde{E}_{F}+264   \tilde{E}_{F}^2+105  \tilde{E}_{F}^3+14   \tilde{E}_{F}^4\right)\bigg]\cos{2\beta} \Bigg\}\Bigg]\Bigg/ \bigg[6   \tilde{E}_{F} (1+  \tilde{E}_{F}) \bigg\{6400  \tilde{E}_{F}^4 \left(2+3  \tilde{E}_{F}+  \tilde{E}_{F}^2\right)^2\\
&+{\tilde{B}}^4 \left(52+28  \tilde{E}_{F}+15   \tilde{E}_{F}^2+19  \tilde{E}_{F}^3 +8   \tilde{E}_{F}^{4}\right)+80 {\tilde{B}}^2   \tilde{E}_{F}^2 \left(56+136  \tilde{E}_{F}+124   \tilde{E}_{F}^2+53   \tilde{E}_{F}^3+9  \tilde{E}_{F}^4\right)\bigg\}\bigg].
\end{aligned}
\end{equation}
The out-of-plane SC is calculated as
\begin{equation}\label{outofplanesc}
\tilde{\nu}_{zz}=\frac{\pi ^2 \left[-80  \tilde{E}_{F}^3 \left(4+7 \tilde{E}_{F}+3  \tilde{E}_{F}^2\right)+{\tilde{B}}^2 \left(8+8  \tilde{E}_{F}-3  \tilde{E}_{F}^2+ \tilde{E}_{F}^3\right)\right]}{6  \tilde{E}_{F} (1+ \tilde{E}_{F}) \left[{\tilde{B}}^2 \left(2- \tilde{E}_{F}+ \tilde{E}_{F}^2\right)+80  \tilde{E}_{F}^2 \left(2+3  \tilde{E}_{F}+ \tilde{E}_{F}^2\right)\right]}.
\end{equation}\\
The coefficient of the planar Nernst effect is given by
\begin{equation}\label{pnc}
\centering
\begin{aligned}
\tilde{\nu}_{xy}^{(2)} & =  {5{\tilde{B}}^2 \pi ^2 \Bigg[{\tilde{B}}^2 \left(-16+4  \tilde{E}_{F}+7   \tilde{E}_{F}^2\right)+16  \tilde{E}_{F} \left(96+272   \tilde{E}_{F}+264   \tilde{E}_{F}^2+105   \tilde{E}_{F}^3+14   \tilde{E}_{F}^4\right)\Bigg] \sin2\beta}\bigg/\\
& 6 \Bigg[6400  \tilde{E}_{F}^4 \left(2+3  \tilde{E}_{F}+ \tilde{E}_{F}^2\right)^2+{\tilde{B}}^4 \left(52+28   \tilde{E}_{F}+15   \tilde{E}_{F}^2+19  \tilde{E}_{F}^3+8   \tilde{E}_{F}^4\right)\\
&+80 {\tilde{B}}^2   \tilde{E}_{F}^2 \left(56+136  \tilde{E}_{F}+124  \tilde{E}_{F}^2+53   \tilde{E}_{F}^3+9  \tilde{E}_{F}^4\right)\Bigg].
\end{aligned}
\end{equation}
For the other case of ${\mathbf{B}}\parallel {\hat{\mathbf{z}}}$, the longitudinal SC is calculated as
\begin{equation}\label{longitudinalsc}
\tilde{\nu}_{zz}=\frac{\pi ^2 \left[-80  \tilde{E}_{F}^3 \left(4+7  \tilde{E}_{F}+3  \tilde{E}_{F}^2\right)+{\tilde{B}}^2 \left(104+184  \tilde{E}_{F}+81  \tilde{E}_{F}^2+8 \tilde{E}_{F}^3\right)\right]}{6  \tilde{E}_{F} (1+ \tilde{E}_{F}) \left[80 \tilde{E}_{F}^2 \left(2+3  \tilde{E}_{F}+ \tilde{E}_{F}^2\right)+{\tilde{B}}^2 \left(26+27  \tilde{E}_{F}+8  \tilde{E}_{F}^2\right)\right]}.
\end{equation}
The perpendicular SC is obtained as
\begin{equation}\label{perpendicularsc}
\tilde{\nu}_{xx}=\nu_{yy}=\frac{\pi ^2 \left[-80 \tilde{E}_{F}^3 \left(4+7  \tilde{E}_{F}+3  \tilde{E}_{F}^2\right)+{\tilde{B}}^2 \left(8+8  \tilde{E}_{F}-3  \tilde{E}_{F}^2+ \tilde{E}_{F}^3\right)\right]}{6  \tilde{E}_{F} (1+ \tilde{E}_{F}) \left[{\tilde{B}}^2 \left(2- \tilde{E}_{F}+ \tilde{E}_{F}^2\right)+80  \tilde{E}_{F}^2 \left(2+3  \tilde{E}_{F}+ \tilde{E}_{F}^2\right)\right]}.
\end{equation}
\end{widetext}


\begin{thebibliography}{55}
\bibitem{spintronics1}
S. A.Wolf, D. D. Awschalom, R. A. Buhrman, J. M. Daughton,
S. Von Molnar, M. L. Roukes, A. Y. Chtchelkanova, and D. M.
Treger, Science {\bf294}, 1488 (2001).

\bibitem{spintronics2}
R. Winkler, \textit{Spin-Orbit Coupling Effects in Two-dimensional
Electron and Hole Systems} (Springer, Berlin/Heidelberg, 2003).

\bibitem{spintronics3}
I. Zutic, J. Fabian, and S. Das Sarma, Rev. Mod. Phys.
{\bf76}, 323 (2004).

\bibitem{spintronics4}
S. D. Bader and S. S. P. Parkin, Annu. Rev. Condens.
Matter Phys {\bf1}, 71 (2010).

\bibitem{rashba1}
E. I. Rashba, Sov. Phys. Solid State {\bf2}, 1109 (1960).
 
\bibitem{rashba2}
Y. A. Bychkov and E. I. Rashba, J. Phys. C {\bf17}, 6039
(1984).

\bibitem{she1}
J. E. Hirsch, Phys. Rev. Lett. {\bf83}, 1834 (1999).

\bibitem{she2}
S. Zhang, Phys. Rev. Lett. {\bf85}, 393 (2000).

\bibitem{she3}
J. Sinova, D. Culcer, Q. Niu, N. A. Sinitsyn, T. Jungwirth, and
A. H. MacDonald, Phys. Rev. Lett. {\bf92}, 126603 (2004).
 
\bibitem{she4}
Y. K. Kato, R. C. Myers, A. C. Gossard, and D. D. Awschalom,
Science {\bf306}, 1910 (2004).

\bibitem{she5}
B. A. Bernevig and S. C. Zhang, Phys. Rev. Lett. {\bf96}, 106802
(2006).

\bibitem{she6}
J. Sinova, S. O. Valenzuela, J. Wunderlich, C. H. Back, and T.
Jungwirth, Rev. Mod. Phys. {\bf87}, 1213 (2015).

\bibitem{TI}
B. A. Bernevig and T. L. Hughes, \textit{Topological insulators and topological superconductors} (Princeton university press, 2013).

\bibitem{sot}
K. Tsutsui and S. Murakami, Phys. Rev. B {\bf86}, 115201
(2012).

\bibitem{sge}
S. D. Ganichev, E. L. Ivchenko, V. V. Belkov, S. A. Tarasenko,
M. Sollinger, D. Weiss, W. Wegscheider, and W. Prettl, Nature
{\bf417}, 153 (2002).

\bibitem{spinorbitronics}
A. Manchon, H. C. Koo, J. Nitta, S. M. Frolov and R. A. Duine,
Nature Mater {\bf14}, 871–882 (2015).

\bibitem{Bi/Ag}
C. R. Ast, J. Henk, A. Ernst, L. Moreschini, M. C. Falub, D.
Pacile, P. Bruno, K. Kern, and M. Grioni, Phys. Rev. Lett. 98,
186807 (2007).

\bibitem{Bi2se31}
M. Bianchi, D. Guan, S. Bao, J.Mi, B. B. Iversen, P. D. C. King,
and P. Hofmann, Nat. Commun. {\bf1}, 128 (2010).

\bibitem{Bi2se32}
P. D. C. King, R. C. Hatch, M. Bianchi, R. Ovsyannikov, C.
Lupulescu, G. Landolt, B. Slomski, J. H. Dil, D. Guan, J. L.
Mi, E. D. L. Rienks, J. Fink, A. Lindblad, S. Svensson, S. Bao,
G. Balakrishnan, B. B. Iversen, J. Osterwalder, W. Eberhardt,
F. Baumberger, and Ph. Hofmann, Phys. Rev. Lett. {\bf107}, 096802
(2011).

\bibitem{giant1}
K. Ishizaka, M. S. Bahramy, H. Murakawa, M. Sakano, T.
Shimojima, T. Sonobe, K. Koizumi, S. Shin, H. Miyahara, A.
Kimura, K. Miyamoto, T. Okuda, H. Namatame, M. Taniguchi,
R. Arita, N. Nagaosa, K. Kobayashi, Y. Murakami, R. Kumai,
Y. Kaneko, Y. Onose, and Y. Tokura, Nat. Mater. {\bf10}, 521
(2011).

\bibitem{giant2}
M. S. Bahramy, R. Arita, and N. Nagaosa, Phys. Rev. B {\bf84},
041202(R) (2011).

\bibitem{giant3}
S. V. Eremeev, I. A. Nechaev, Yu. M. Koroteev, P. M.
Echenique, and E. V. Chulkov, Phys. Rev. Lett. {\bf108}, 246802
(2012).

\bibitem{giant4}
G. Landolt, S. V. Eremeev, Y. M. Koroteev, B. Slomski, S.
Muff, T. Neupert, M. Kobayashi, V. N. Strocov, T. Schmitt,
Z. S. Aliev, M. B. Babanly, I. R. Amiraslanov, E. V. Chulkov, J.
Osterwalder, and J. H. Dil, Phys. Rev. Lett. {\bf109}, 116403 (2012).

\bibitem{giant5}
M. Sakano, M. S. Bahramy, A. Katayama, T. Shimojima,
H. Murakawa, Y. Kaneko, W. Malaeb, S. Shin, K. Ono, H.
Kumigashira, R. Arita, N. Nagaosa, H. Y. Hwang, Y. Tokura,
and K. Ishizaka, Phys. Rev. Lett. {\bf110}, 107204 (2013).

\bibitem{B20}
J. Kang and J. Zang, Phys. Rev. B {\bf91}, 134401 (2015).

\bibitem{LiPd}
V. P. Mineev and Y. Yoshioka, Phys. Rev. B {\bf81}, 094525
(2010).

\bibitem{zhou}
S.-X. Wang, H.-R. Chang and J. Zhou, Phys. Rev. 
B {\bf96}, 115204 (2017).

\bibitem{symmetryLiPd}
K.-W. Lee and W. E. Pickett, Phys. Rev. B {\bf72}, 174505
(2005).

\bibitem{symmetry1}
K. V. Samokhin, Phys. Rev. B {\bf78}, 144511 (2008).

\bibitem{berry}
M. V. Berry, Quantal phase factors accompanying adiabatic
changes, Proc. R. Soc. London, Ser. A {\bf392}, 45 (1984).

\bibitem{sonu}
S. Verma, T. Biswas, and T. K. Ghosh,
Phys. Rev. B {\bf100}, 045201 (2019).

\bibitem{xiao}
D. Xiao, M.-C. Chang, and Q. Niu, Rev. Mod. Phys. {\bf82},
1959 (2010).

\bibitem{AHE1}
T. Jungwirth, Q. Niu, and A. H. MacDonald, Phys. Rev.
Lett. {\bf88}, 207208 (2002).
 
\bibitem{AHE2}
A. A. Burkov, Phys.
Rev. Lett. {\bf113}, 187202 (2014).

\bibitem{kamal}
K. Das and A. Agarwal,
Phys. Rev. B {\bf99}, 085405 (2019).

\bibitem{ANE1}
G. Sharma, P. Goswami, and S. Tewari, 
Phys. Rev. B {\bf93}, 035116 (2016).

\bibitem{ANE2}
G. Sharma, C. Moore, S. Saha, S. Tewari, Phys. Rev. B {\bf96}, 195119
(2017).

\bibitem{ANE3}
G. Sharma and S. Tewari,
Phys. Rev. B {\bf100}, 195113 (2019).
 
\bibitem{CME1}
D. T. Son and N. Yamamoto,
Phys. Rev. Lett., {\bf109}, 181602 (2012).

\bibitem{CME2}
A. A. Zyuzin and A. A. Burkov, 
Phys. Rev. B {\bf86}, 115133 (2012).

\bibitem{chiral1}
S. Adler, Phys. Rev. {\bf177}, 2426 (1969).

\bibitem{chiral2}
J. S. Bell and R. A. Jackiw, Nuovo Cimento A {\bf60}, 47 (1969).

\bibitem{chiral3}
H. B. Nielsen and M. Ninomiya, Phys. Lett. B {\bf105}, 219
(1981).

\bibitem{sasaki}
K.-S. Kim, H.-J. Kim, and M. Sasaki, Phys. Rev. B {\bf89},
195137 (2014).

\bibitem{1}
H. J. Kim, K. S. Kim, J. F. Wang, M. Sasaki, N. Satoh,
A. Ohnishi, M. Kitaura, M. Yang, and L. Li, Phys. Rev. Lett.
{\bf111}, 246603 (2013).

\bibitem{2}
Q. Li, D. E. Kharzeev, C. Zhang, Y. Huang, I. Pletikosic,
A. V. Fedorov, R. D. Zhong, J. A. Schneeloch, G. D. Gu,
and T. Valla, Nat. Phys. {\bf12},
550 (2016).

\bibitem{3}
X. C. Huang, X. Huang, L. Zhao, Y. Long, P. Wang, D. Chen, Z. Yang, H. Liang, M. Xue, H. Weng, Z. Fang, X. Dai, and G. Chen, Phys. Rev. X {\bf5}, 031023 (2015).

\bibitem{4}
J. Xiong, S. K. Kushwaha, T. Liang, J.W. Krizan, M.
Hirschberger,W.Wang, R. J. Cava, and N. P. Ong, Science {\bf350}, 413 (2015).

\bibitem{5}
C. Z. Li, L. X. Wang, H.W. Liu, J. Wang, Z. M. Liao, and
D. P. Yu, Nat. Commun. {\bf6}, 10137 (2015).

\bibitem{6}
C. Zhang, E. Zhang, W. Wang, Y. Liu, Z.-G. Chen, S. Lu, S. Liang, J. Cao, X. Yuan, L. Tang, Q. Li, C. Zhou, T. Gu, Y. Wu, J. Zou and F. Xiu, Nat. Commun. {\bf8}, 13741 (2017).

\bibitem{7}
H. Li, H. T. He, H. Z. Lu, H. C. Zhang, H. C. Liu, R. Ma,
Z. Y. Fan, S. Q. Shen, and J. N. Wang, Nat. Commun. {\bf7},
10301 (2016).

\bibitem{8}
F. Arnold, C. Shekhar, SC. Wu, Y. Sun, R. D. dos Reis, N. Kumar, M. Naumann, M. O. Ajeesh, M. Schmidt, A. G. Grushin, J. H. Bardarson, M. Baenitz, D. Sokolov, H. Borrmann, M. Nicklas, C. Felser, E. Hassinger and B. Yan , Nat.
Commun. {\bf7}, 11615 (2016).

\bibitem{9}
H. Wang, C.-K. Li, H. Liu, J. Yan, J. Wang, J. Liu, Z. Lin, Y. Li, Y. Wang, L. Li, D. Mandrus, X. C. Xie, J. Feng, and J. Wang, Phys. Rev. B {\bf93}, 165127 (2016).

\bibitem{lu}
X. Dai, Z. Z. Du, and H.-Z. Lu, Phys. Rev. Lett. {\bf119}, 
166601 (2017).

\bibitem{nandy} 
S. Nandy, G. Sharma, A. Taraphder, and S. Tewari, Phys.
Rev. Lett. {\bf119}, 176804 (2017).

\bibitem{burkov} 
A. A. Burkov, 
Phys. Rev. B {\bf96}, 041110 (2017).

\bibitem{kamal2}
K. Das and A. Agarwal,
Phys. Rev. B {\bf100}, 085406 (2019). 

\bibitem{pne1 ferro} 
C. T. Bui and F. Rivadulla, Phys. Rev. B {\bf90}, 100403 (2014).

\bibitem{pne2 ferro} 
C. T. Bui, C. A. C. Garcia, N. T. Tu, M. Tanaka, and
P. N. Hai, Journal of Applied Physics {\bf123}, 175102 (2018).


\bibitem{niu2}
D. Xiao, Y. Yao, Z. Fang, and Q. Niu, Phys. Rev. Lett.
{\bf97}, 026603 (2006).


\bibitem{OMM1}
M.-C. Chang and Q. Niu, Phys. Rev. B {\bf53}, 7010 (1996).

\bibitem{OMM2}
D. Xiao, W. Yao, and Q. Niu, Phys. Rev. Lett. {\bf99}, 236809
(2007).
  
\bibitem{moore}
T. Morimoto, S. Zhong, J. Orenstein, and J. E. Moore,
Phys. Rev. B {\bf94}, 245121 (2016).
 
\bibitem{mdos}
D. Xiao, J. Shi, and Q. Niu, Phys. Rev. Lett. {\bf95}, 137204
(2005).


\bibitem{prlniu}
Y. Gao, S. A. Yang, and Q. Niu, Phys. Rev. Lett. {\bf112}, 166601
(2014).

\bibitem{potential}
Y. Gao, S. A. Yang, and Q. Niu, Phys. Rev. B {\bf91}, 214405
(2015).

\bibitem{shift}
Y. Gao, S. A. Yang, and Q. Niu, Phys. Rev. B {\bf95}, 165135
(2017).

\bibitem{ashcroft}
N. Ashcroft and N. Mermin, Solid State Physics, HRW
international editions (Holt, Rinehart and Winston, 1976).
 
\bibitem{jacobini}
C. Jacoboni,\textit{Theory of electron transport in semiconduc-
tors: a pathway from elementary physics to nonequilibrium
green functions} (Springer, Berlin, 2010).

\bibitem{negmrniu}
H. Zhou, C. Xiao, and Q. Niu, Phys. Rev. B {\bf100}, 041406 (2019).

\bibitem{halperin}
N. R. Cooper, B. I. Halperin, and I. M. Ruzin, Phys. Rev.
B 55, 2344 (1997).

\bibitem{jackson}
J. D. Jackson, \textit{Classical Electrodynamics} (John Wiley $\&$
Sons, New York, 1999), 3rd ed.

\end{thebibliography}
\end{document}